\documentclass[runningheads]{llncs}
\usepackage{amsmath,amsfonts}
\usepackage{algorithmic}
\usepackage{graphicx}
\usepackage{textcomp}
\usepackage{xcolor}
\usepackage{booktabs}
\usepackage[hidelinks]{hyperref}
\usepackage{tikz}
\usepackage{float}
\usepackage{multirow}
\usepackage{booktabs}
\usepackage{comment}
\usepackage{subfigure}
\usepackage{amsmath}
\usepackage{xspace}
\usepackage{balance}
\usepackage{todonotes}
\usepackage{caption} 
\usepackage{newtxmath}
\usepackage{dblfloatfix}
\usepackage{gensymb}
\usepackage{lipsum} 
\usepackage[]{mdframed}
\usepackage{tcolorbox}

\newcommand{\attack}{$Lens Attack$\xspace}
\newcommand\cbox[1]{\begin{tcolorbox}#1\end{tcolorbox}}

\begin{document}
\title{Optical Lens Attack on Deep Learning Based Monocular Depth Estimation}
%

\author{Ce Zhou\inst{1} \and
Qiben Yan\inst{1}\thanks{Corresponding author.} \and
Daniel Kent\inst{1} \and
Guangjing Wang\inst{1} \and
Ziqi Zhang\inst{2} \and
Hayder Radha\inst{1}
}
\authorrunning{Zhou et al.}
%
\institute{Michigan State University, East Lansing MI 48823, USA \and
Peking University, Beijing 100871, China}
%
\maketitle              
\begin{abstract}
  Monocular Depth Estimation (MDE) plays a crucial role in vision-based Autonomous Driving (AD) systems. It utilizes a single-camera image to determine the depth of objects, facilitating driving decisions such as braking a few meters in front of a detected obstacle or changing lanes to avoid collision. In this paper, we investigate the security risks associated with monocular vision-based depth estimation algorithms utilized by AD systems. By exploiting the vulnerabilities of MDE and the principles of optical lenses, we introduce \attack, a physical attack that involves strategically placing optical lenses on the camera of an autonomous vehicle to manipulate the perceived object depths. \attack encompasses two attack formats: concave lens attack and convex lens attack, each utilizing different optical lenses to induce false depth perception. We begin by constructing a mathematical model of our attack, incorporating various attack parameters. Subsequently, we simulate the attack and evaluate its real-world performance in driving scenarios to demonstrate its effect on state-of-the-art MDE models. The results highlight the significant impact of \attack on the accuracy of depth estimation in AD systems.

\end{abstract}
%
%
%
\section{Introduction}


Tracking and maintaining the distance to surrounding obstacles is a key function of Autonomous Driving (AD) systems' perception modules, without which the AD systems cannot operate safely and reliably. There is a wide range of solutions for performing this task, e.g., through direct measurements using radar or Lidar~\cite{piotrowsky2019enabling,li2020lidar}, or using stereoscopic 3D imaging to reconstruct a dense depth map of the scene~\cite{mayer2016large,chang2018pyramid,tay2019aanet}. 

Camera is one of the most important sensors in AD systems, as seen in vehicles from Tesla~\cite{Autopilot}, Uber~\cite{Uber} and Waymo~\cite{Waymo}, and these vehicles rely on the computer vision technology for AD tasks~\cite{zhou2023comprehensive}. Researchers have developed an advanced technology that enables monocular cameras to estimate scene depth~\cite{monodepth2,wong2020targeted,casser2019depth}. Solutions using monocular cameras would therefore reduce the number of sensors required, saving valuable space, weight, and cost. Despite the challenges in estimating depth information using monocular cameras, recent deep-learning based methods have reached performance levels comparable to stereo 3D depth estimation techniques. 
 
Existing security studies propose different attack methods towards cameras to disrupt various AD tasks, such as object detection and classification~\cite{eykholt2018robust,man2020ghostimage,nassi2020phantom}, lane detection~\cite{jing2021too,sato2021dirty},  traffic light detection~\cite{yan2022rolling}, and vision-based depth estimation~\cite{zhou2022doublestar,zhang2020adversarial,wong2020targeted}.  
To compromise 3D stereo depth estimation, Zhou et al.~\cite{zhou2022doublestar} propose a long-range stereo depth estimation attack that injects fake obstacle depth by projecting pure light from two complementary light sources. 
For monocular depth estimation (MDE) algorithms, 
Zhang et al.~\cite{zhang2020adversarial} and Wong et al.~\cite{wong2020targeted} present white-box attacks that use imperceptible additive adversarial perturbations to alter the depth estimation results in the digital world, while a black-box attack is introduced by Daimo et al. in \cite{daimo2021black}. However, these invisible perturbations are ineffective in the physical world due to the impacts of environmental variables. Therefore, Yamanaka et al.~\cite{yamanaka2020adversarial} and Cheng et al.~\cite{cheng2022physical} create visible adversarial patches. These patches deceive depth estimation algorithms into estimating a false depth for the regions where the patterns are placed in the physical world. However, human drivers can easily detect the patches. Moreover, patches are scene-sensitive, and may not work well in dynamic environments. As opposed to existing physical attacks, we propose a universal black-box attack, \attack, that enables a new type of robust physical attack using optical lenses. 

\attack exploits the inherent vulnerability of MDE, i.e., a small alternation of the object size in an image could result in a corresponding change in depth. Our attack utilizes the optical lens to change the formed object size on the image sensor. Specifically, by attaching a tiny attack lens in the near front (e.g., $5cm$) of the car camera, the sensed object size will be altered, which affects the depth estimation results. 

There are two major challenges in realizing \attack. (i) ``How to design the attacks that can induce various false depth predictions?'' (ii) ``How to mathematically calculate the induced depth and gain control over depth estimation?'' 
To address the first challenge, 
we design \attack in two attack formats: concave lens attack and convex lens attack, which can either increase or decrease the object depth. 
To solve the second challenge, we mathematically model our attack using different lenses in various attack scenarios. 

We verify the efficacy of \attack through both simulation and real-world experiments with a prototype autonomous vehicle (AV) against three state-of-the-art MDE algorithms. The results demonstrate that our attack remains effective across a wide range of optical lens parameter configurations. 
We set up a demo website\footnote{\url{https://lensattack.github.io/.}} to show our attack results, attack simulations, and physical attack video demos.

The main contributions of this work are summarized as follows:
\begin{itemize}
\item We propose a novel universal physical attack, \attack, on MDE algorithms that utilizes optical lenses. 
\item By investigating the vulnerability of the MDE, we propose the concave and convex lens attacks and mathematically model them in different attack scenarios. 
\item We show potential attack consequences in the simulation and physical world on three state-of-the-art MDE models. We evaluate the attacks on a smartphone camera and an AV in real-world experiments to demonstrate that our attack is effective with various optical lens parameter settings. The concave lens attack results in an average error rate of 11.48\% in estimated depths, whereas the convex lens attack leads to a 29.84\% average error rate. 


\end{itemize}

\section{Background}

In this section, we briefly introduce the preliminary background knowledge of \attack, including the optical principles for optical lenses and monocular vision based depth estimation.

\subsection{Optical Principles for Lenses}\label{sec:background}
An optical lens, typically made of transparent materials such as glasses, is employed to produce an image through the concentration of light rays emanating from an object~\cite{Britannica}. This is accomplished by exploiting the phenomenon of refraction, which arises when light passes from one medium (such as air) to another (the lens). As a result, refraction occurs both upon entering the lens and upon exiting it back into the air. Optical lenses have wide-ranging applications, including but not limited to, eyeglasses, magnifiers, projection condensers, signal lights, viewfinders, cameras, and more.

An optical lens typically has a circular shape and possesses two polished surfaces, which can either be concave or convex. There are different types of lenses based on the curvature of the two opposite surfaces. Regarding the prevalence and availability, we mainly focus on the lens whose two surfaces are both concave or convex, which is called \emph{double concave} or \emph{double convex} lens. For simplicity, we will refer to them as ``concave'' and ``convex'' lenses. 

\begin{figure}[t]
    \centering    
    \subfigure[]{\includegraphics[width=0.24\textwidth]{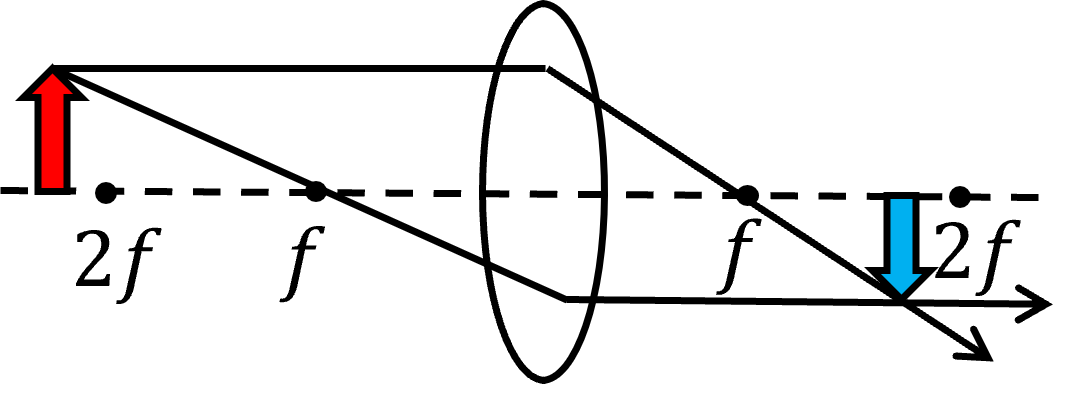}} 
    \subfigure[]{\includegraphics[width=0.25\textwidth]{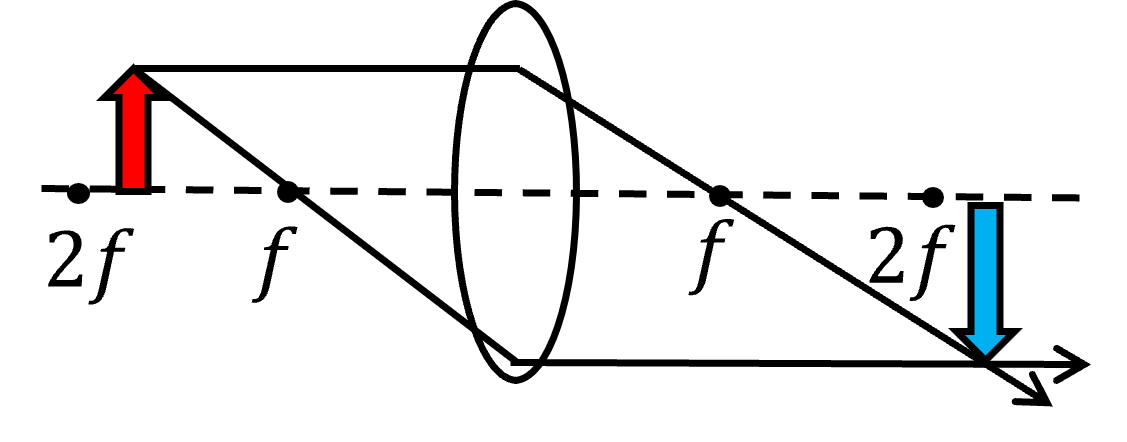}} 
    \subfigure[]{\includegraphics[width=0.21\textwidth]{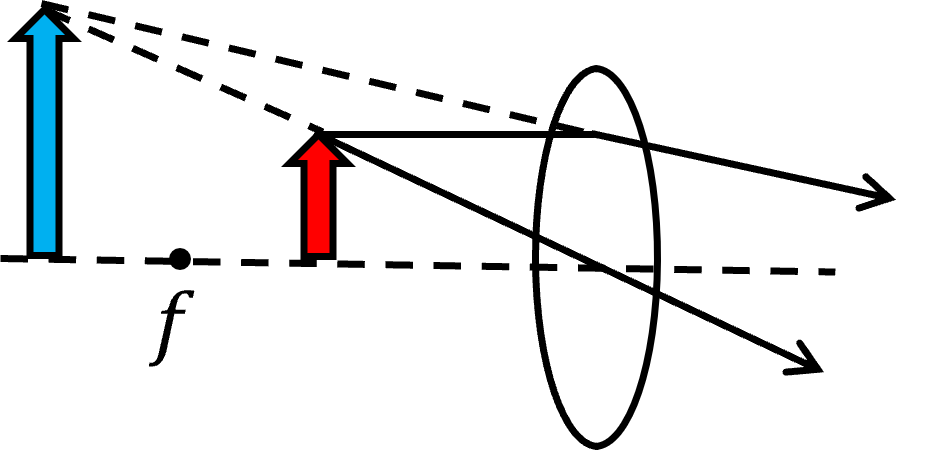}}   
    \subfigure[]{\includegraphics[width=0.21\textwidth]{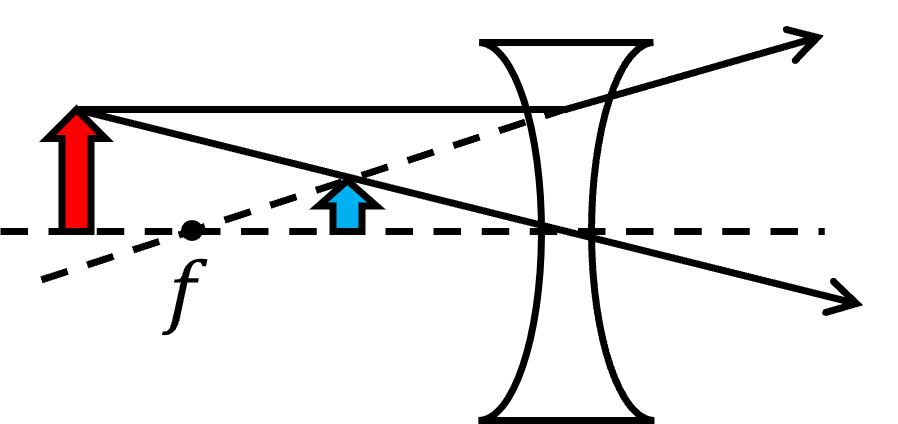}} 
    \caption{Ray diagrams of concave and convex lenses. (a)(b)(c) show the images formed by a convex lens, and (d) displays the images formed by a concave lens. The object is shown in red arrow and its corresponding image formed by the attack lens is shown in blue. $f$ represents the focal length.}
    \label{fig:background_oplens}
\end{figure}

A focal point, also known as a principal focus (denoted as $f$ in Fig. \ref{fig:background_oplens}), is the point at which parallel rays can be made to converge or appear to diverge~\cite{Britannica,Focus}. A lens has two focal points, one on each side, so light can pass through it in either direction. The distance from the center of the lens to the focal point is the \emph{focal length}.

Fig. \ref{fig:background_oplens} shows the visual image of the object formed by a convex lens and a concave lens due to ray refractions. For the convex lens, the size of the images formed can vary significantly compared to the object, depending on the focal length of the lens $f$ and the distance between the lens and the object $d_o$. There exist three possible cases: (Case 1) When $d_o \geq 2f$, it forms a real, inverted and smaller image where the distance between the lens and the image is $f\leq d_i\leq 2f$. (Case 2) When $f < d_o < 2f$, it creates a real, inverted and larger image at $d_i> 2f$. (Case 3) When $ 0< d_o < f$, it produces a virtual, upright, and larger image behind the object on the same side of the lens. On the other hand, the images formed by a concave lens are always virtual, upright, and smaller between the object and the lens regardless of the object's position.

The relationship between $f$, $d_o$ and $d_i$ can be written as: 
\begin{equation}\label{equ1}
    \frac{1}{f}=\frac{1}{d_o}+\frac{1}{d_i},
\end{equation}
where $f$ is a positive number if it is a convex lens, otherwise a negative number, $d_o$ is always a positive number, and $d_i$ is a positive number if the lens is on the opposite side of the object, otherwise a negative number.

The magnification of the formed image is:
\begin{equation}\label{equ2}
    m=-\frac{d_i}{d_o}.
\end{equation}
If $m$ is a positive number, the image is upright. 

Based on Eqs.~(\ref{equ1}) and (\ref{equ2}), in Case (3) with images formed by a convex lens, a larger absolute value of focal length results in a smaller image. In other cases, a larger absolute value of focal length leads to a larger image of an object, whereas a smaller absolute value of focal length forms a smaller image.



\subsection{Monocular Depth Estimation}\label{section:MonocularDepthEstimation}
The objective of MDE is to determine each pixel's depth value from a single 2D RGB image.  MDE has become popular in the study of robotic tasks~\cite{datar2023learning,datar2023toward}. Previous studies use supervised training methods to estimate the depth from a single image~\cite{liu2015deep,eigen2015predicting,dijk2019neural,yin2019enforcing}. The lack of high-quality depth maps, however, has spurred the adoption of unsupervised/self-supervised learning. Instead of using crowd-sourced data~\cite{chen2016single} for training, the recently proposed methods use stereo-pairs~\cite{garg2016unsupervised,godard2017unsupervised,pillai2019superdepth} or monocular video sequences~\cite{zhou2017unsupervised,wang2018learning,casser2019depth,yin2018geonet,guizilini20203d}. Video-based algorithms for depth estimation typically provide depth values in an unknown scaled format, whereas stereo-based methods can predict depth in metric units if the baseline between the cameras is known~\cite{wong2020targeted}.

Monodepth2~\cite{monodepth2} uses both stereo and video-based methods and leverages reprojection losses to eliminate potential occlusions. However, reprojection losses from stereo-based self-supervision typically have multiple local minima which restrict the network learning, and further lead to the limited quality of the predicted depth map. Depth Hints~\cite{watson2019self} enhances an existing loss function to better guide a network to learn weights. 
However, these algorithms are hard to be deployed on edge devices due to its heavy computation. Lite-mono~\cite{zhang2023lite} is a lightweight but effective model, which significantly reduces the number of trainable parameters, by employing a hybrid CNN and Transformer architecture. In this paper, we aim to attack these three cutting-edge methods to demonstrate the broad impact of our proposed attack.

A pinhole camera is a simple model for approximating the monocular camera imaging process. Although the actual imaging is much more complicated than the pinhole imaging model, the pinhole imaging model is very convenient to apply mathematically, and the approximation to imaging is often acceptable~\cite{Pinholecameramodel}. 

Based on the pinhole imaging model (more details can be found in Appendix~\ref{sec:Pinhole Camera Model}), we learn that the depth of the same object is inversely proportional to its object size in the image. 
A change in the size of an object in the image will result in a corresponding change in depth. Therefore, for monocular camera imaging, as the distance between the object and the camera increases, its size becomes smaller on an image, and vice versa. 

MDE algorithms are vulnerable due to the lack of sufficient physical indications for scene depth in a monocular image alone. Current MDE algorithms appear to rely on implicit knowledge learned from the training dataset (e.g., color, location, or shadows) rather than physical cues~\cite{dijk2019neural,yamanaka2020adversarial}). Because they rely heavily on non-depth elements in the provided image, they are vulnerable to attacks that tamper with the images.

\section{Threat Model and Attack Scenario}\label{sec:threat_model}

In this section, we introduce our threat model and present the attack scenarios.

\subsection{Threat Model}

The attacker's goal is to disrupt the regular operations of an AV by alternating the MDE results using the optical lens and triggering unintended system behaviors. 

We consider an AV relying on the monocular camera as the major source for depth estimation, such as Tesla vehicles with Full Self-Driving features and Active Safety Features~\cite{Autopilot_and_full_self}.
The attacker tries to change the depth information of the objects, e.g., vehicles or pedestrians, by applying optical lenses within a near distance (e.g., $5cm$) to the target camera. As a result, this can cause the victim AV to crash and potentially lead to a severe car accident, such as colliding with the vehicle in front. 


Our attack is a \emph{black-box attack} against general monocular vision-based depth estimation algorithms in AD systems. The attacker is assumed to have no prior knowledge of the depth estimation algorithms used in AD systems. We assume the attacker does not have access to the camera images. We also assume that an attacker has very limited physical access to the hardware or firmware of the victim. 
For example, if the victim's vehicle is parked in a public area, the attacker can get close to the vehicle and install the attacking equipment, e.g., using fixed suction cups to absorb the attack device on the camera or the car body depending on the outlook of the camera. 

The attacker may need to use a 3D-designed lens holder
to hold the attack lens to ensure the stability of the attack lens during driving. Typically, the attacker would design 
the 3D lens holder in a compact size and with a transparent color or a color similar to that of the car body. This design approach is crucial for achieving a relatively stealthy attack. The attacker could also design the attack lenses with a proper size to make the attacks more accurate and stealthy. The implementation details of the attack device on a real vehicle are illustrated in Appendix~\ref{app:manufactured}.


\subsection{Attack Scenario}

\begin{figure}[t]
\centering
	\includegraphics[width=1\columnwidth]{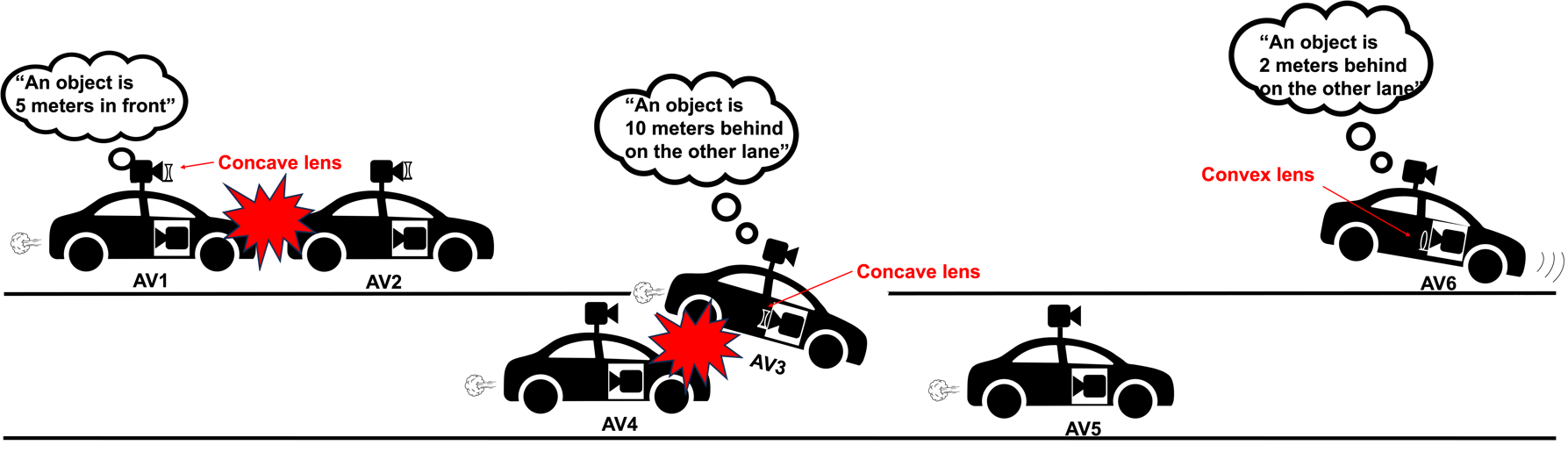}
	\caption{\attack scenarios. AV1 to AV4 illustrate the concave lens attacks, and AV5 and AV6 showcase a convex lens attack.}
	\label{fig:Op_lens_attack}
\end{figure}

We present attack scenarios for normal driving and lane changing in Fig.~\ref{fig:Op_lens_attack}. 
 A concave lens is attached to the front camera of AV1, which extends the estimated depth for AV2 beyond its real distance. This false depth estimation could lead to a collision, posing a significant threat to human lives, or cause a hard brake that may injure passengers if other sensors are present and sensor fusion is employed. Even if the AV1 detects the proximity of the front car and attempts to brake, the distance between the AV1 and AV2 is smaller than the ``safety braking distance'', which is insufficient to prevent a collision. Besides, the AV2 may make a sudden lane change to avoid collision with the front vehicle, but this could potentially result in severe traffic jams or car accidents if the following vehicle on the other lane does not have sufficient time to react. Additionally, when a concave lens is applied to the side camera of AV3 during the lane changing, the depth estimation of AV4 by AV3 is farther away than its actual depth. As a result, AV3 shifts lanes accordingly, which could cause a potential collision with AV4 or hard braking of AV4. Similarly, a convex lens on the side camera of AV6 results in an estimated depth of AV5 being nearer than its actual depth. This may lead to delayed lane changing or, in certain situations, force AV5 to stop and wait to change lanes, potentially causing a traffic jam.

\section{Optical Lens Attack} \label{sec:attack_design}



\attack exploits the vulnerabilities in the monocular depth perception. The basic idea is to apply the attack lenses in front of the camera lens to form a combination lens system, which will enlarge or reduce the size of the sensed images. As shown in Fig.~\ref{fig:att_design}, we propose four different lens combinations based on convex and concave lens attacks. As a result, it causes the depth change of the objects. The parameters used in this section along with their respective meanings are summarized in Table~\ref{tab:symbol}.

\begin{table}[t]
\centering
\caption{List of parameters and their meanings}
\label{tab:symbol}
\resizebox{0.7\columnwidth}{!}{%
\begin{tabular}{|l|l|l|}
\hline
\textbf{Symbol} &
  \textbf{Description} &
  \textbf{Sign} \\ \hline
$f$ &
  The focal length of the attack lens &
  \begin{tabular}[c]{@{}l@{}}Concave lens: negative;\\ Convex lens: positive\end{tabular} \\ \hline
$d_{o1}$ &
  \begin{tabular}[c]{@{}l@{}}The distance between the object\\ and the attack lens\end{tabular} &
  Positive \\ \hline
$d_{i1}$ &
  \begin{tabular}[c]{@{}l@{}}The distance between the attack lens\\ and the formed lens image\end{tabular} &
  \begin{tabular}[c]{@{}l@{}}Virtual image: positive;\\ real image: negative\end{tabular} \\ \hline
$m_1$ &
  \begin{tabular}[c]{@{}l@{}}Magnification of the formed attack lens\\ image compared to the real object size\end{tabular} &
  \begin{tabular}[c]{@{}l@{}}Virtual image: positive;\\ real image: negative\end{tabular} \\ \hline
$d_{i2}$ &
  \begin{tabular}[c]{@{}l@{}}The distance between the attack lens\\ and the formed camera image\end{tabular} &
  \begin{tabular}[c]{@{}l@{}}Virtual image: positive;\\ real image: negative\end{tabular} \\ \hline
$m_2$ &
  \begin{tabular}[c]{@{}l@{}}Magnification of the formed camera\\ image compared to the attack lens image\end{tabular} &
  \begin{tabular}[c]{@{}l@{}}Virtual image: positive;\\ real image: negative\end{tabular} \\ \hline
$f_c$ &
  The focal length of the camera lens &
  Positive \\ \hline
$d_b$ &
  \begin{tabular}[c]{@{}l@{}}The distance between the attack lens\\ and the camera lens\end{tabular} &
  Positive \\ \hline
$m_{total}$ &
  \begin{tabular}[c]{@{}l@{}}The total magnification of the formed camera\\ image compared to the real object size\end{tabular} &
  \begin{tabular}[c]{@{}l@{}}Virtual image: positive;\\ real image: negative\end{tabular} \\ \hline
$m_{ori}$ &
  \begin{tabular}[c]{@{}l@{}}Magnification of the formed camera image\\ compared to the real object size without \\ applying the attack lens\end{tabular} &
  \begin{tabular}[c]{@{}l@{}}Virtual image: positive;\\ real image: negative\end{tabular} \\ \hline
\end{tabular}%
}
\end{table}


\subsection{Concave Lens Attack}
\begin{figure*}[t]
    \centering
    \subfigure[]{\includegraphics[width=0.21\textwidth]{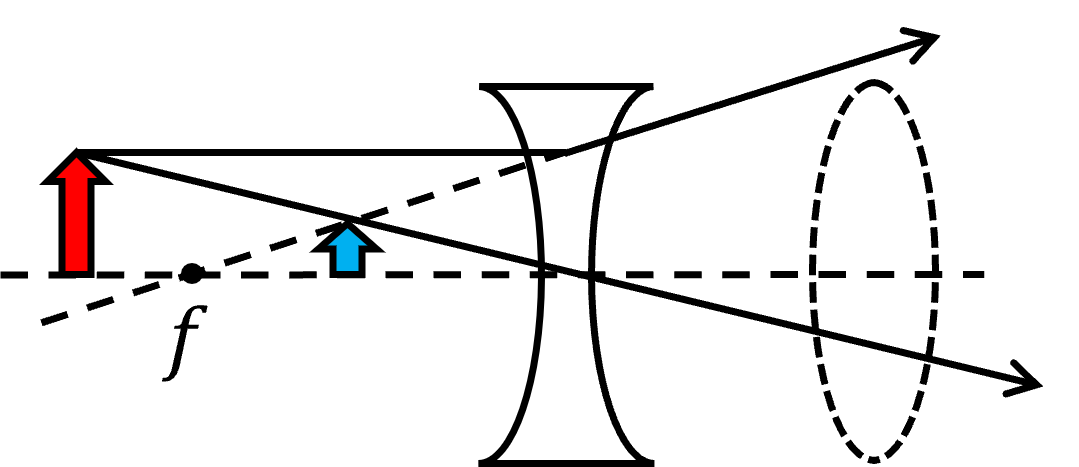}} 
    \subfigure[]{\includegraphics[width=0.23\textwidth]{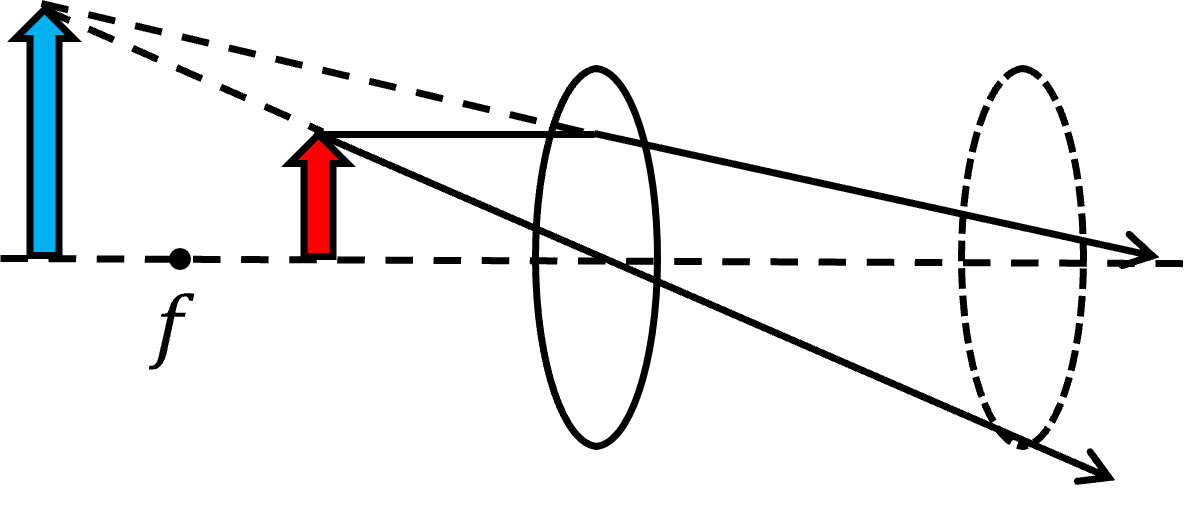}} 
    \subfigure[]{\includegraphics[width=0.25\textwidth]{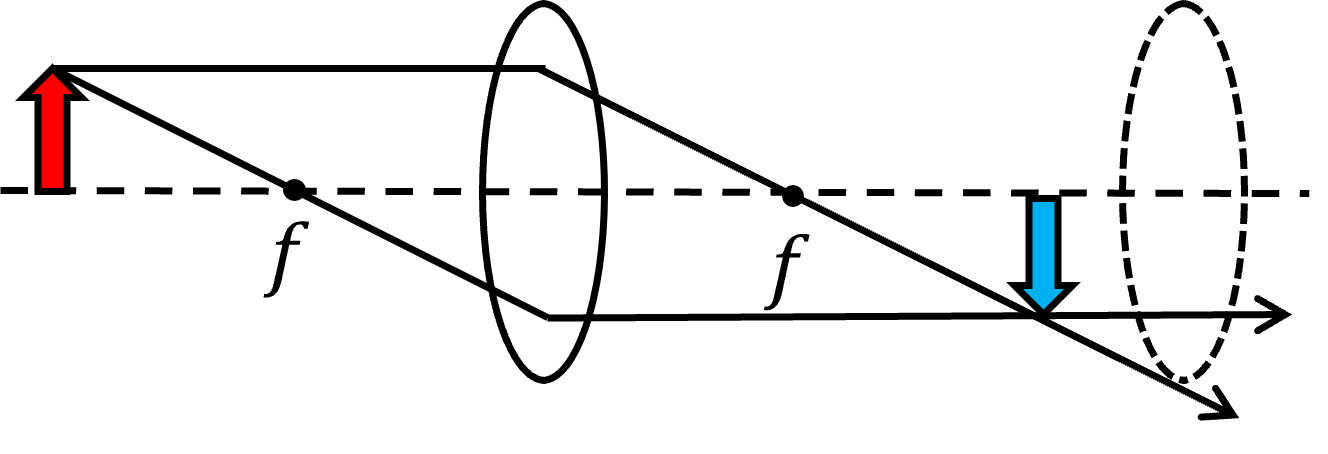}}
    \subfigure[]{\includegraphics[width=0.25\textwidth]{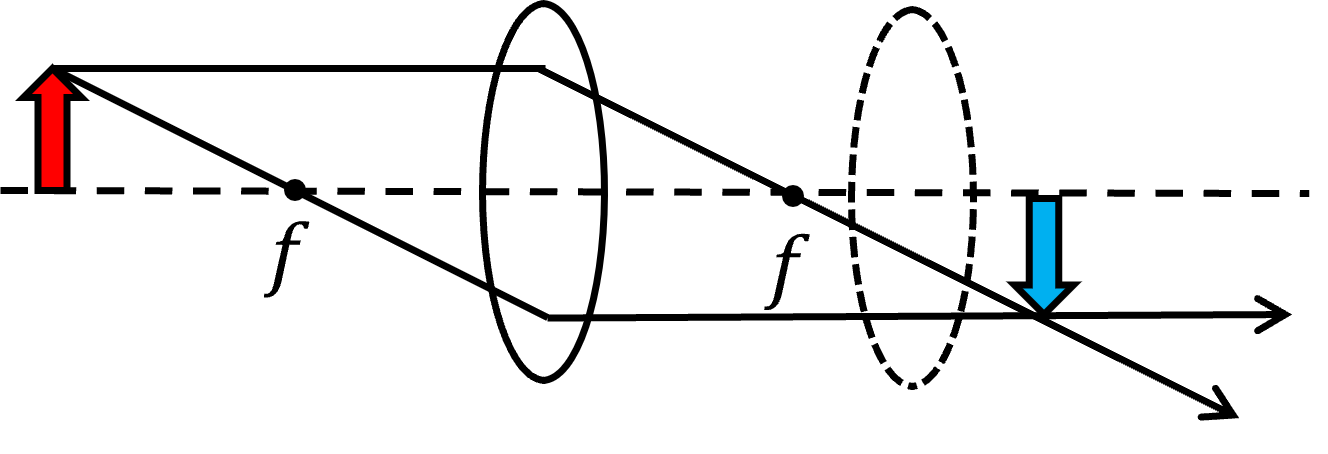}}
    \caption{Ray diagrams for (a) concave lens attack, and (b) first (c) second (d) third attack scenarios in convex lens attack. The left lens is the attack lens and the right dotted one stands for the monocular camera lens. The object is shown in red arrow and its corresponding image formed by the attack lens is shown in blue. $f$ represents the focal length of the attack lens.}
    \label{fig:att_design}
\end{figure*}

As discussed in Section~\ref{sec:background}, only one type of image is formed by the concave lens. The lens combination is shown in Fig.~\ref{fig:att_design}(a).
Based on Eq.~(\ref{equ1}), we have:
\begin{equation}\label{equ}
    \frac{1}{f}=\frac{1}{d_{o1}}+\frac{1}{d_{i1}},
\end{equation}
where $d_{o1}$ stands for the distance between the object and the attack lens, and $d_{i1}$ denotes the distance between the attack lens and the image.
Therefore, $d_{i1}$ can be written as:
\begin{equation}\label{equ}
    d_{i1}=-\frac{d_{o1} f}{d_{o1}-f}.
\end{equation}

Based on Eq.~(\ref{equ2}), we have the magnification $m_1$ for the attack lens as:
\begin{equation}\label{equ}
    m_1=-\frac{d_{i1}}{d_{o1}}=-\frac{f}{d_{o1}-f}.
\end{equation}

Similarly, for the camera lens, we have the distance between the lens and the image $d_{i2}$ and magnification $m_2$ as:
\begin{equation}\label{equ}
    d_{i2}=-\frac{(|d_{i1}|+d_b) f_c}{(|d_{i1}|+d_b)-f_c},
\end{equation}
and
\begin{equation}\label{equ}
    m_2=-\frac{d_{i2}}{d_{o2}}=-\frac{f_c}{(|d_{i1}|+d_b)-f_c},
\end{equation}
where $f_c$ is the focal length of the camera lens and $d_b$ is the distance between the attack lens and the camera lens. $d_b$ is always a positive number.

Thus, the total magnification $m_{total}$ can be expressed as:
\begin{equation}\label{equ:concave}
    m_{total}=m_1m_2=\frac{ff_c}{(d_{o1}-f)(|\frac{d_{o1}f}{d_{o1}-f}|+d_b-f_c)}.
\end{equation}

In the vehicle's camera, the focal length is rather small. The object's distance (i.e., the distance between the object and the camera) is usually much larger than twice the focal length, so the image is always formed at the focal point (i.e., the location of the image sensor) regardless of the object's distance. Therefore, the size of the image formed on the image sensor is inversely proportional to the object's distance. The larger the image is formed, the smaller the object's distance can be. Since the original magnification $m_{ori}$ is:
\begin{equation}\label{equ:magnification}
    m_{ori}=-\frac{d_{i}}{d_{o}}=-\frac{f_c}{d_{o1}+d_b-f_c},
\end{equation}
the manipulated depth becomes $|m_{ori}/m_{total}|$ of the original depth.

\subsection{Convex Lens Attack}
There are three attack cases based on the different distances between the object and the lens (Figs.~\ref{fig:att_design}(b)(c)(d)). If $0<d_{o1}<f$ (\textit{as in the first attack scenario}), we can place the camera anywhere we desire. It has the same mathematical analysis as in the concave lens attack. On the other hand, if $d_{o1} > f$, $d_b$ may be further away than the image formed by the attack lens, indicating $d_b\geq d_{i1}$ (\textit{as in the second attack scenario}). Or, it could be closer than the image's location formed by the attack lens, indicating $d_b\leq d_{i1}$ (\textit{as in the third attack scenario}). 

Following the same derivation procedure as the convex lens attack, we have the same expression of $d_{i1}$ and $m_1$ for the attack lens. For the second attack scenario ($d_b\geq d_{i1}$), $d_{i2}$ and $m_2$ for the camera lens can be expressed as:
\begin{equation}\label{equ}
    d_{i2}=-\frac{(d_b-|d_{i1}|) f_c}{(d_b-|d_{i1}|)-f_c},
\end{equation}
\begin{equation}\label{equ}
    m_2=-\frac{d_{i2}}{d_{o2}}=-\frac{f_c}{(d_b-|d_{i1}|)-f_c}.
\end{equation}
Thus, the total magnification $m_{total}$ is:
\begin{equation}\label{equ}
    m_{total}=m_1m_2=\frac{ff_c}{(d_{o1}-f)(d_b-|\frac{d_{o1}f}{d_{o1}-f}|-f_c)}.
\end{equation}
For the third attack scenario ($d_b\leq d_{i1}$), $d_{i2}$ and $m_2$ for the camera lens can be expressed as:
\begin{equation}\label{equ}
    d_{i2}=-\frac{(|d_{i1}|-d_b) f_c}{(|d_{i1}|-d_b)-f_c},
\end{equation}
\begin{equation}\label{equ}
    m_2=-\frac{d_{i2}}{d_{o2}}=-\frac{f_c}{(|d_{i1}|-d_b)-f_c}.
\end{equation}
Thus, the total magnification $m_{total}$ is: 
\begin{equation}\label{equ:convex}
    m_{total}=m_1m_2=\frac{ff_c}{(d_{o1}-f)(|\frac{d_{o1}f}{d_{o1}-f}|-d_b-f_c)}.
\end{equation}

In theory, all the attack scenarios should be feasible. However, due to the constraints in physical attacks, only the concave lens attack and the third scenario of the convex lens attack would lead to a successful attack. More details are presented in Sections~\ref{sec:evl} and~\ref{sec:discussion}.

\subsection{Coverage of the Attack Lens}

We design two types of attack lens coverage in \attack: full and partial. If the attack lens covers the entire image, the depth of the entire image will be altered. Conversely, if only part of the image is covered by the attack lens, the depth of that portion of the image is expected to change while the rest of the image remains unaffected. However, the resulting depth measurements may not be as accurate as calculated due to the camera's focusing feature.

A photographic lens that does not possess adjustable focus capability is referred to as a fixed-focus lens~\cite{fixed_focus_wiki}.
Typically, advanced driver-assistance systems (ADAS), drones, and AD cameras are fixed-focus~\cite{wittpahl2018realistic}, because it is best for handling high-vibration environments~\cite{fixed_focus}. When we add an optical lens in front of the camera, it will separate the whole image into the in-lens area and out-of-lens area if the lens is in the camera view. The focal length of the in-lens area will be the combination of the focal length from the optical focal length and the victim camera lens. The focal lens of the out-of-lens area is still the same as the benign one. Since the in-lens and out-of-lens have different focal lengths, blur usually will be added to the in-lens area due to the depth of field (DOF) effect. Note that DOF is defined as the distance between the closest and farthest objects in an image captured with a camera that is in acceptable sharp focus~\cite{Depthoffield}. 

On the other hand, the camera with autofocus (AF), a function of a camera to automatically focus on a subject (e.g., smartphone camera), can also be affected by DOF in our attack. Due to the focal length difference caused by the addition of the attack lens, the camera will either focus on the in-lens or out-of-lens areas, leading to blurriness in the part that is not in focus. Blur in the image can affect the depth estimation results and contribute to the average depth error rate in the later physical experiments. 

\section{Evaluation} \label{sec:evl}

To demonstrate the impact of \attack, we begin by simulating it in the digital world and then evaluate its performance in the physical world. 

\subsection{Attack Setup}\label{sec:attack_setup}

\subsubsection{Target Models}
To evaluate the accuracy and effectiveness of our attack, we consider three state-of-the-art MDE algorithms,  Monodepth2~\cite{monodepth2},  Depth Hints~\cite{watson2019self}, and Lite-mono~\cite{zhang2023lite} as the attack targets. They are stereo and video-based, stereo-based, and video-based algorithms, respectively. The selection is based on their representativeness, timeliness, and popularity. The output of the Monodepth2 can be a depth map or disparity map, whereas the output of Depth Hints and Lite-mono is only a disparity map. The disparity map can be converted into the depth map if it is trained on the stereo pairs and the baseline and focal length are known. The relation between the disparity and depth is as follows: 
\begin{equation}
    disparity = \frac{(baseline*focal\:length)}{depth}.
\end{equation}
Note that we only examine the attack performance on Monodepth2 when the evaluation requires the real depth value.

\subsubsection{Evaluation Metrics}
We adopt Attack Distortion Rate (ADR) and Attack Error Rate (AER) as the evaluation metrics to show the performance of the attack in both the digital world and the physical world. The ADR and AER are defined as follows:
\begin{equation}
    ADR = \frac{|attacked\:depth-benign\:depth|}{benign\:depth},
\end{equation}
and 
\begin{equation}
    AER = \frac{|attacked\:depth-target\:depth|}{target\:depth}.
\end{equation}
A higher ADR implies a greater depth difference caused by the attack. Conversely, a lower AER indicates a smaller deviation from the target or expected depth value, indicating a higher attack success rate. 



\subsubsection{Dataset in Simulation}
All three target models are trained and evaluated on the KITTI dataset~\cite{geiger2013vision}. Therefore, for the attack simulation, we use images from the KITTI semantic split~\cite{abu2018augmented}. We investigate both full and partial coverage of the attack lens. In the partial coverage scenarios, the attack area is circular, matching the shape of the physical attack lens. We apply different ratios to enlarge or reduce the size of the attack area and then compare the attack results on the disparity map.

\begin{figure*}[t]
    \centering    
    \subfigure[Full view of the attack]{\includegraphics[width=0.4\textwidth]{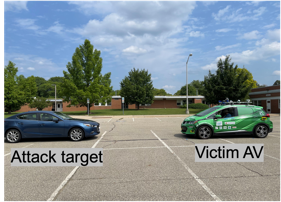}}    
    \subfigure[Data collection on AV]{\includegraphics[width=0.4\textwidth]{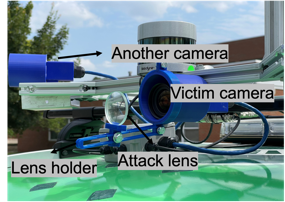}}
    \subfigure[3D-printed lens holder]{\includegraphics[width=0.4\textwidth]{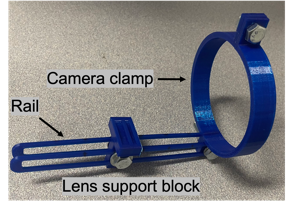}\label{fig:5c}} 
    \subfigure[Data collection in driving scenario]{\includegraphics[width=0.4\textwidth]{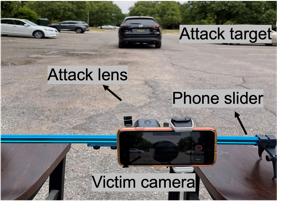}}
    \caption{Experimental setup for the AV and real-world driving.}
    \label{fig:experimental_setup}
\end{figure*}

\subsubsection{Physical Experimental Setup}
The human eye has a focal length of approximately $22mm$~\cite{cambridgeincolour}. Car cameras usually have a similar focal length as the human eye's. A longer focal length leads to a narrower field of view. As for AD vehicles, a shorter focal length provides a broader field of view. Besides, we find that the commonly available optical lenses in the market without particular order or design are usually within the focal length of $50mm$ to $500mm$.

In the physical experiments, we test our attack on an AV camera and a smartphone camera. The experimental setup is shown in Fig.~\ref{fig:experimental_setup}. The victim AV is a 2017 Chevy Bolt Electric Vehicle with three FLIR BFS-PGE-31S4C cameras mounted on the car's roof, which will be called ``AV'' for simplicity. The focal length of the FLIR camera with lens is approximately $23mm$. To collect the attack images in real world, we design a lens holder that is 3D printable as shown in Fig.~\ref{fig:5c}. It contains a camera clamp, a rail, and a lens support block. The lens support block can hold the lens and slide on the rail so that we can adjust the distance between the attack lens and the camera lens ($d_b$). The lens holder can be positioned externally to the FLIR camera such that the victim AV can collect images that contain the target vehicle. 

In real driving scenarios (see Fig.~\ref{fig:experimental_setup}(d)), we collect images of driving scenes using iPhone 12 Pro Max, whose focal length is $26mm$ \cite{dxomark}. We will call it ``iPhone'' for the rest of the content. Since the focal lengths of the FLIR camera and the iPhone camera are almost the same, we will take $26mm$ for the expected attack depth calculation. Two sets of concave and convex lenses~\cite{AmlongCrystal} are used in the physical attacks with focal lengths of $20cm$, $30cm$, and $50cm$ in each set. To compare the attack results, we use various values of $f$, $d_b$, and $d_o$. For the real-world physical attack, we have the attack setup inside the victim's vehicle. We use a phone slider to hold the attack lens so that we can remotely control the position of the attack lens using a wireless controller, allowing for a more flexible attack. We have also integrated YOLO v8 with the MDE algorithm in the real-world driving scenario. Physical attack video demos are available on our website.

\subsection{Attack Simulation in Digital World} 

The goal of \attack is to modify the target object size on an image such that the estimated depth can be manipulated. We consider the lens can be applied to either the entire image or a portion of it. We simulate our attack in three attack scenarios: full image cropping (or full image enlarging), partial image enlarging, and partial image shrinking. Note that full image shrinking is just the reverse case of full image enlarging, so we do not consider it separately. 

\begin{figure}
\centering
	\includegraphics[width=0.7\columnwidth]{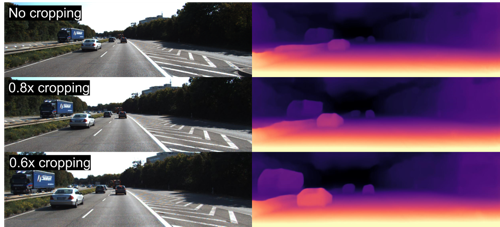}
	\caption{Full image cropping with 0.8x and 0.6x cropping ratios.}
	\label{fig:cropping}
\end{figure}


\begin{figure}
\centering
	\includegraphics[width=0.7\columnwidth]{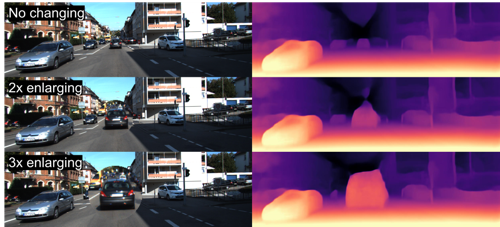}
	\caption{Partial image enlarging with 2x and 3x enlarging ratios.}
	\label{fig:enlarging}
\end{figure}

\begin{figure}
\centering
	\includegraphics[width=0.7\columnwidth]{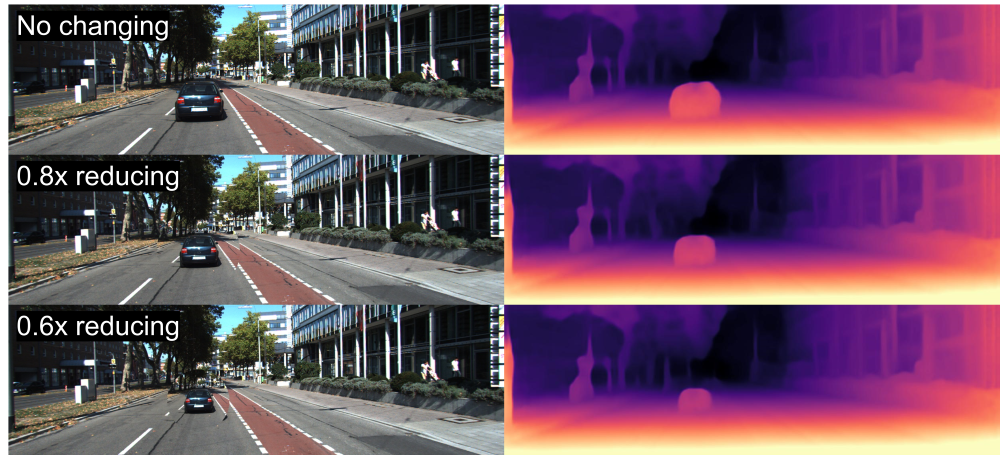}
	\caption{Partial image reducing with 0.8x and 0.6x reducing ratios.}
	\label{fig:reducing}
\end{figure}

\begin{table*}
\centering
\caption{Masked mean object depth and the corresponding ADR in the attack simulation with different image modifications }
\label{tab:simulation_results}
\resizebox{1\textwidth}{!}{%
\begin{tabular}{|cc|c|ccc|ccc|}
\hline
\multicolumn{2}{|c|}{\multirow{2}{*}{Image Modification}} &
  \multirow{2}{*}{\begin{tabular}[c]{@{}c@{}}Masked Object Mean Depth Value \\ in Monodepth2 (meters)\end{tabular}} &
  \multicolumn{3}{c|}{Masked Object Mean Disparity Value} &
  \multicolumn{3}{c|}{ADR} \\ \cline{4-9} 
\multicolumn{2}{|c|}{} &
   &
  \multicolumn{1}{c|}{Monodepth2} &
  \multicolumn{1}{c|}{Depth Hints} &
  Lite-Mono &
  \multicolumn{1}{c|}{Monodepth2} &
  \multicolumn{1}{c|}{Depth Hints} &
  Lite-Mono \\ \hline
\multicolumn{1}{|c|}{\multirow{3}{*}{Cropping}} &
  Bengin &
  21.08 &
  \multicolumn{1}{c|}{0.28} &
  \multicolumn{1}{c|}{0.31} &
  1.89 &
  \multicolumn{1}{c|}{-} &
  \multicolumn{1}{c|}{-} &
  - \\ \cline{2-9} 
\multicolumn{1}{|c|}{} &
  0.8x &
  16.44 &
  \multicolumn{1}{c|}{0.36} &
  \multicolumn{1}{c|}{0.37} &
  2.19 &
  \multicolumn{1}{c|}{28.6\%} &
  \multicolumn{1}{c|}{19.3\%} &
  15.9\% \\ \cline{2-9} 
\multicolumn{1}{|c|}{} &
  0.6x &
  14.00 &
  \multicolumn{1}{c|}{0.45} &
  \multicolumn{1}{c|}{0.46} &
  2.66 &
  \multicolumn{1}{c|}{60.7\%} &
  \multicolumn{1}{c|}{48.4\%} &
  40.7\% \\ \hline
\multicolumn{1}{|c|}{\multirow{3}{*}{Enlarging}} &
  Bengin &
  24.66 &
  \multicolumn{1}{c|}{0.23} &
  \multicolumn{1}{c|}{0.23} &
  1.47 &
  \multicolumn{1}{c|}{-} &
  \multicolumn{1}{c|}{-} &
  - \\ \cline{2-9} 
\multicolumn{1}{|c|}{} &
  2x &
  17.29 &
  \multicolumn{1}{c|}{0.36} &
  \multicolumn{1}{c|}{0.36} &
  2.18 &
  \multicolumn{1}{c|}{56.5\%} &
  \multicolumn{1}{c|}{56.5\%} &
  48.3\% \\ \cline{2-9} 
\multicolumn{1}{|c|}{} &
  3x &
  11.37 &
  \multicolumn{1}{c|}{0.50} &
  \multicolumn{1}{c|}{0.52} &
  3.03 &
  \multicolumn{1}{c|}{117.4\%} &
  \multicolumn{1}{c|}{126.1\%} &
  106.1\% \\ \hline
\multicolumn{1}{|c|}{\multirow{3}{*}{Shrinking}} &
  Bengin &
  13.97 &
  \multicolumn{1}{c|}{0.44} &
  \multicolumn{1}{c|}{0.46} &
  2.68 &
  \multicolumn{1}{c|}{-} &
  \multicolumn{1}{c|}{-} &
  - \\ \cline{2-9} 
\multicolumn{1}{|c|}{} &
  0.8x &
  14.30 &
  \multicolumn{1}{c|}{0.40} &
  \multicolumn{1}{c|}{0.41} &
  2.42 &
  \multicolumn{1}{c|}{9\%} &
  \multicolumn{1}{c|}{10.9\%} &
  9.7\% \\ \cline{2-9} 
\multicolumn{1}{|c|}{} &
  0.6x &
  14.78 &
  \multicolumn{1}{c|}{0.38} &
  \multicolumn{1}{c|}{0.38} &
  2.24 &
  \multicolumn{1}{c|}{13.6\%} &
  \multicolumn{1}{c|}{17.4\%} &
  16.4\% \\ \hline
\end{tabular}%
}
\end{table*}

To show the concept and feasibility of our attack,
the attack simulation is conducted on all three MDE algorithms without the consideration of blurriness. The attack results of full image cropping on Monodepth2 are shown in Fig.~\ref{fig:cropping}. The simulated images are shown in the left column, and the corresponding disparity maps are displayed in the right column. When the attack lens is applied to the whole image, we can see that the more we crop the image, the more the depth can be altered. Similarly, when the attack lens is applied to the partial image in Figs.~\ref{fig:enlarging} and ~\ref{fig:reducing}, the more we enlarge or shrink the attack area, the more the depth can be changed. 


Meanwhile, we also mask the target object using the object detection results from YOLO v8, and compute its mean depth value in the masked area which represents the average distance of the target object to the camera. In Table~\ref{tab:simulation_results}, we show the mean depth value of masked objects in Monodepth2 in meters, the mean disparity value of masked objects in all target models, and their corresponding ADRs. It can be observed that partially enlarging the target object changes the depth the most. With $3x$ enlarging, the depth of the target vehicle can be reduced by around $13m$ with the corresponding ADR over 100\% for all three depth estimation algorithms. Then, the full image cropping comes second regarding the extent of manipulating the depth of the target vehicle. 
Shrinking the size of the target vehicle can only change the depth up to $1m$ with  $0.6x$ image shrinking. Therefore, the results show that different image modifications can cause the depth of the target object to vary. 


\cbox{To summarize, \attack has caused similar depth distortion effect 
across all three target models, demonstrating the generalization potential of the attack.}


\subsection{Real-World Physical Attack}\label{sec:Real-World Physical Attack}

To further investigate the practicality and generality, we launch \attack in the physical world. We perform both concave lens attacks and convex lens attacks with various object distances, i.e., $6m$, $9m$, and $12m$. 

\subsubsection{Performance of Concave Lens Attacks}

For the concave lens attack, we investigate the effects of $f$ and $d_b$ in altering the target object depth with different $d_{o1}$.
The results in Table~\ref{tab:attack_error_rate_concave} indicate a smaller $f$ leads to a larger change in the object depth. It also shows that smaller $d_b$ results in less impact on the depth. 

\begin{table*}[]
\centering
\caption{The AER of the concave lens attack with the various object distance $d_{o1}$, $f$, and $d_b$.}
\label{tab:attack_error_rate_concave}
\resizebox{\textwidth}{!}{%
\begin{tabular}{|cc|ccccc|ccccc|ccccc|}
\hline
\multicolumn{2}{|c|}{\multirow{2}{*}{Attack Parameters}} &
  \multicolumn{5}{c|}{$d_{o1}$=6m} &
  \multicolumn{5}{c|}{$d_{o1}$=9m} &
  \multicolumn{5}{c|}{$d_{o1}$=12m} \\ \cline{3-17} 
\multicolumn{2}{|c|}{} &
  \multicolumn{1}{c|}{\multirow{2}{*}{\begin{tabular}[c]{@{}c@{}}Expected\\ Depth (meters)\end{tabular}}} &
  \multicolumn{2}{c|}{\begin{tabular}[c]{@{}c@{}}Experimental Depth\\ (meters)\end{tabular}} &
  \multicolumn{2}{c|}{\begin{tabular}[c]{@{}c@{}}AER\\ (\%)\end{tabular}} &
  \multicolumn{1}{c|}{\multirow{2}{*}{\begin{tabular}[c]{@{}c@{}}Expected\\ Depth (meters)\end{tabular}}} &
  \multicolumn{2}{c|}{\begin{tabular}[c]{@{}c@{}}Experimental Depth\\ (meters)\end{tabular}} &
  \multicolumn{2}{c|}{\begin{tabular}[c]{@{}c@{}}AER\\ (\%)\end{tabular}} &
  \multicolumn{1}{c|}{\multirow{2}{*}{\begin{tabular}[c]{@{}c@{}}Expected\\ Depth (meters)\end{tabular}}} &
  \multicolumn{2}{c|}{\begin{tabular}[c]{@{}c@{}}Experimental Depth\\ (meters)\end{tabular}} &
  \multicolumn{2}{c|}{\begin{tabular}[c]{@{}c@{}}AER\\ (\%)\end{tabular}} \\ \cline{1-2} \cline{4-7} \cline{9-12} \cline{14-17} 
\multicolumn{1}{|c|}{$f$} &
  $d_b$ &
  \multicolumn{1}{c|}{} &
  \multicolumn{1}{c|}{AV} &
  \multicolumn{1}{c|}{iPhone} &
  \multicolumn{1}{c|}{AV} &
  iPhone &
  \multicolumn{1}{c|}{} &
  \multicolumn{1}{c|}{AV} &
  \multicolumn{1}{c|}{iPhone} &
  \multicolumn{1}{c|}{AV} &
  iPhone &
  \multicolumn{1}{c|}{} &
  \multicolumn{1}{c|}{AV} &
  \multicolumn{1}{c|}{iPhone} &
  \multicolumn{1}{c|}{AV} &
  iPhone \\ \hline
\multicolumn{1}{|c|}{20cm} &
  2cm &
  \multicolumn{1}{c|}{5.82} &
  \multicolumn{1}{c|}{-} &
  \multicolumn{1}{c|}{7.09} &
  \multicolumn{1}{c|}{-} &
  21.9 &
  \multicolumn{1}{c|}{8.73} &
  \multicolumn{1}{c|}{-} &
  \multicolumn{1}{c|}{7.67} &
  \multicolumn{1}{c|}{-} &
  12.12 &
  \multicolumn{1}{c|}{11.64} &
  \multicolumn{1}{c|}{-} &
  \multicolumn{1}{c|}{10.62} &
  \multicolumn{1}{c|}{-} &
  8.76 \\ \hline
\multicolumn{1}{|c|}{20cm} &
  4cm &
  \multicolumn{1}{c|}{6.42} &
  \multicolumn{1}{c|}{5.85} &
  \multicolumn{1}{c|}{6.85} &
  \multicolumn{1}{c|}{9.10} &
  6.77 &
  \multicolumn{1}{c|}{9.63} &
  \multicolumn{1}{c|}{9.82} &
  \multicolumn{1}{c|}{10.54} &
  \multicolumn{1}{c|}{1.97} &
  9.44 &
  \multicolumn{1}{c|}{12.84} &
  \multicolumn{1}{c|}{11.82} &
  \multicolumn{1}{c|}{12.46} &
  \multicolumn{1}{c|}{7.91} &
  2.92 \\ \hline
\multicolumn{1}{|c|}{20cm} &
  8cm &
  \multicolumn{1}{c|}{7.61} &
  \multicolumn{1}{c|}{7.73} &
  \multicolumn{1}{c|}{6.00} &
  \multicolumn{1}{c|}{1.59} &
  21.13 &
  \multicolumn{1}{c|}{11.42} &
  \multicolumn{1}{c|}{10.89} &
  \multicolumn{1}{c|}{12.65} &
  \multicolumn{1}{c|}{4.62} &
  10.80 &
  \multicolumn{1}{c|}{15.23} &
  \multicolumn{1}{c|}{12.86} &
  \multicolumn{1}{c|}{13.58} &
  \multicolumn{1}{c|}{15.55} &
  10.81 \\ \hline
\multicolumn{1}{|c|}{20cm} &
  12cm &
  \multicolumn{1}{c|}{8.78} &
  \multicolumn{1}{c|}{5.13} &
  \multicolumn{1}{c|}{6.89} &
  \multicolumn{1}{c|}{41.56} &
  21.46 &
  \multicolumn{1}{c|}{13.19} &
  \multicolumn{1}{c|}{10.56} &
  \multicolumn{1}{c|}{13.62} &
  \multicolumn{1}{c|}{19.93} &
  3.26 &
  \multicolumn{1}{c|}{17.60} &
  \multicolumn{1}{c|}{12.26} &
  \multicolumn{1}{c|}{16.73} &
  \multicolumn{1}{c|}{30.32} &
  4.90 \\ \hline
\multicolumn{1}{|c|}{30cm} &
  2cm &
  \multicolumn{1}{c|}{5.88} &
  \multicolumn{1}{c|}{-} &
  \multicolumn{1}{c|}{7.05} &
  \multicolumn{1}{c|}{-} &
  19.82 &
  \multicolumn{1}{c|}{8.82} &
  \multicolumn{1}{c|}{-} &
  \multicolumn{1}{c|}{8.37} &
  \multicolumn{1}{c|}{-} &
  5.09 &
  \multicolumn{1}{c|}{11.76} &
  \multicolumn{1}{c|}{-} &
  \multicolumn{1}{c|}{12.28} &
  \multicolumn{1}{c|}{-} &
  4.40 \\ \hline
\multicolumn{1}{|c|}{30cm} &
  4cm &
  \multicolumn{1}{c|}{6.28} &
  \multicolumn{1}{c|}{6.37} &
  \multicolumn{1}{c|}{7.08} &
  \multicolumn{1}{c|}{1.42} &
  12.78 &
  \multicolumn{1}{c|}{9.42} &
  \multicolumn{1}{c|}{8.87} &
  \multicolumn{1}{c|}{10.54} &
  \multicolumn{1}{c|}{5.79} &
  11.93 &
  \multicolumn{1}{c|}{12.56} &
  \multicolumn{1}{c|}{12.02} &
  \multicolumn{1}{c|}{12.45} &
  \multicolumn{1}{c|}{4.29} &
  0.90 \\ \hline
\multicolumn{1}{|c|}{30cm} &
  8cm &
  \multicolumn{1}{c|}{7.07} &
  \multicolumn{1}{c|}{6.39} &
  \multicolumn{1}{c|}{7.51} &
  \multicolumn{1}{c|}{9.64} &
  6.22 &
  \multicolumn{1}{c|}{10.61} &
  \multicolumn{1}{c|}{8.78} &
  \multicolumn{1}{c|}{11.91} &
  \multicolumn{1}{c|}{17.23} &
  12.23 &
  \multicolumn{1}{c|}{14.15} &
  \multicolumn{1}{c|}{12.02} &
  \multicolumn{1}{c|}{14.65} &
  \multicolumn{1}{c|}{15.06} &
  3.52 \\ \hline
\multicolumn{1}{|c|}{30cm} &
  12cm &
  \multicolumn{1}{c|}{7.85} &
  \multicolumn{1}{c|}{5.10} &
  \multicolumn{1}{c|}{7.78} &
  \multicolumn{1}{c|}{35.04} &
  0.97 &
  \multicolumn{1}{c|}{11.79} &
  \multicolumn{1}{c|}{9.64} &
  \multicolumn{1}{c|}{11.57} &
  \multicolumn{1}{c|}{18.25} &
  1.87 &
  \multicolumn{1}{c|}{15.73} &
  \multicolumn{1}{c|}{13.07} &
  \multicolumn{1}{c|}{15.90} &
  \multicolumn{1}{c|}{16.92} &
  1.04 \\ \hline
\multicolumn{1}{|c|}{50cm} &
  2cm &
  \multicolumn{1}{c|}{5.93} &
  \multicolumn{1}{c|}{-} &
  \multicolumn{1}{c|}{7.15} &
  \multicolumn{1}{c|}{-} &
  20.68 &
  \multicolumn{1}{c|}{8.89} &
  \multicolumn{1}{c|}{-} &
  \multicolumn{1}{c|}{7.21} &
  \multicolumn{1}{c|}{-} &
  18.94 &
  \multicolumn{1}{c|}{11.86} &
  \multicolumn{1}{c|}{-} &
  \multicolumn{1}{c|}{11.44} &
  \multicolumn{1}{c|}{-} &
  3.53 \\ \hline
\multicolumn{1}{|c|}{50cm} &
  4cm &
  \multicolumn{1}{c|}{6.17} &
  \multicolumn{1}{c|}{6.38} &
  \multicolumn{1}{c|}{7.02} &
  \multicolumn{1}{c|}{3.45} &
  13.87 &
  \multicolumn{1}{c|}{9.25} &
  \multicolumn{1}{c|}{8.68} &
  \multicolumn{1}{c|}{10.35} &
  \multicolumn{1}{c|}{6.16} &
  11.82 &
  \multicolumn{1}{c|}{12.34} &
  \multicolumn{1}{c|}{11.39} &
  \multicolumn{1}{c|}{12.56} &
  \multicolumn{1}{c|}{7.67} &
  1.82 \\ \hline
\multicolumn{1}{|c|}{50cm} &
  8cm &
  \multicolumn{1}{c|}{6.64} &
  \multicolumn{1}{c|}{5.82} &
  \multicolumn{1}{c|}{7.71} &
  \multicolumn{1}{c|}{12.42} &
  16.09 &
  \multicolumn{1}{c|}{9.97} &
  \multicolumn{1}{c|}{8.75} &
  \multicolumn{1}{c|}{11.11} &
  \multicolumn{1}{c|}{12.17} &
  11.50 &
  \multicolumn{1}{c|}{13.29} &
  \multicolumn{1}{c|}{11.54} &
  \multicolumn{1}{c|}{13.91} &
  \multicolumn{1}{c|}{13.15} &
  4.66 \\ \hline
\multicolumn{1}{|c|}{50cm} &
  12cm &
  \multicolumn{1}{c|}{7.11} &
  \multicolumn{1}{c|}{5.46} &
  \multicolumn{1}{c|}{6.9} &
  \multicolumn{1}{c|}{23.24} &
  2.97 &
  \multicolumn{1}{c|}{10.67} &
  \multicolumn{1}{c|}{8.97} &
  \multicolumn{1}{c|}{10.32} &
  \multicolumn{1}{c|}{15.99} &
  3.28 &
  \multicolumn{1}{c|}{14.24} &
  \multicolumn{1}{c|}{10.64} &
  \multicolumn{1}{c|}{14.49} &
  \multicolumn{1}{c|}{25.25} &
  1.76 \\ \hline
\multicolumn{1}{|c|}{None} &
  None &
  \multicolumn{1}{c|}{6} &
  \multicolumn{1}{c|}{6.5} &
  \multicolumn{1}{c|}{6.89} &
  \multicolumn{1}{c|}{8.41} &
  14.89 &
  \multicolumn{1}{c|}{9} &
  \multicolumn{1}{c|}{8.60} &
  \multicolumn{1}{c|}{9.94} &
  \multicolumn{1}{c|}{4.49} &
  10.40 &
  \multicolumn{1}{c|}{12} &
  \multicolumn{1}{c|}{11.67} &
  \multicolumn{1}{c|}{12.67} &
  \multicolumn{1}{c|}{2.78} &
  5.59 \\ \hline
\multicolumn{2}{|c|}{\textbf{Average AER}} &
  \multicolumn{3}{c|}{} &
  \multicolumn{1}{c|}{\textbf{15.27}} &
  \textbf{13.72} &
  \multicolumn{3}{c|}{\textbf{}} &
  \multicolumn{1}{c|}{\textbf{11.35}} &
  \textbf{9.36} &
  \multicolumn{3}{c|}{\textbf{}} &
  \multicolumn{1}{c|}{\textbf{15.12}} &
  \textbf{4.09} \\ \hline
\end{tabular}%
}
\end{table*}

To further investigate the AER in the physical world, we calculate the expected depths in meters based on Eqs.~(\ref{equ:concave}) and (\ref{equ:magnification}) and compare them with the experimental values in Table~\ref{tab:attack_error_rate_concave}. The last row shows the prediction accuracy of the Monodepth2 on the benign image. Note that since the depth estimation algorithm is not perfect, the physical experiments introduce some measuring errors, and the camera focusing features, we consider the AER of the depth prediction lower than 15\% as accurate. 

When $d_b=2cm$, the attack lens can fully cover the image on the iPhone camera. However, the attack lens size is always smaller than the FLIR camera lens, so the attack lens we use cannot fully cover the image on AV. We measure the full image attack on iPhone in our experiments. Note that, with the larger size of the attack lens, we are still able to launch the full image attack on AV. The average AER is around 11\% regardless of the focal lengths, which means that the experimental depth value closely matches the expected depth value.

\begin{figure*}[]
\centering
	\includegraphics[width=1\textwidth]{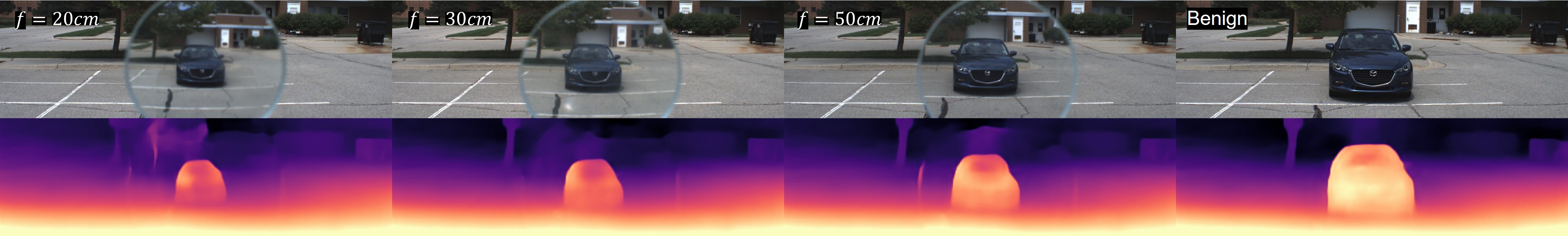}
	\caption{The concave lens attack on an AV with $d_{o1}=9m$, $d_b=12cm$ and different $f$ values.}
	\label{fig:cc_overview_9m_f_12cm}
\end{figure*}

When $d_b$ becomes larger than $4cm$, the attack lens is always in the camera view, which forms the scenario of the partial image attack. We can observe that the AER increases up to around 40\% on AV and 30\% on iPhone. Based on the average AER listed in the last row, we can conclude that our attack works similarly for both AV and iPhone. The overall average AER on both AV and iPhone is 11.48\%. It is also noticeable that the smaller $f$ introduces a higher AER, which is mainly caused by the blur in the in-lens or out-of-lens area due to the effect of DOF. For example, we show the collected images and their corresponding disparity map of concave lens attacks in Fig.~\ref{fig:cc_overview_9m_f_12cm} have the same $d_{o1}=9m$ and $d_b=12cm$ with different $f$ values. 

\vspace{-20pt}
\subsubsection{Performance of Convex Lens Attacks}

Regarding the convex lens attack, we first investigate the first attack scenario. To realize the attack, we need to ensure $d_{o1} < f$. However, the focal length of commonly available convex lenses in the market is usually less than $1m$. Therefore, the first attack scenario is unsuitable for real-world driving because the expected object distance is too short in the attack. 
\begin{figure*}[]
\centering
	\includegraphics[width=1\textwidth]{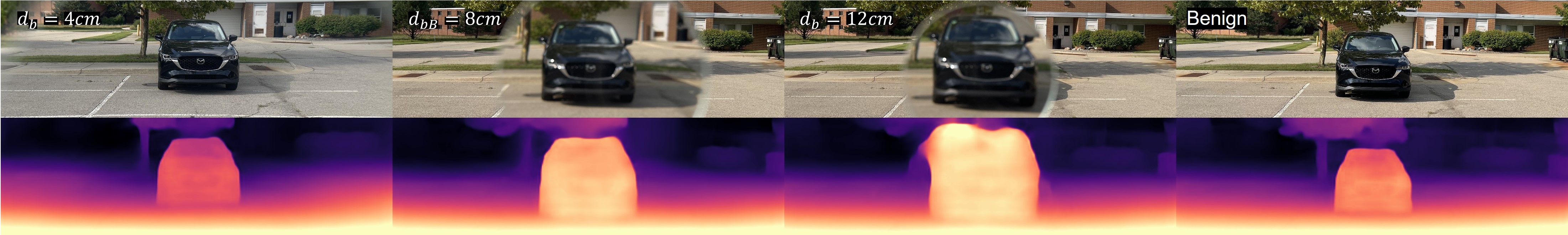}
	\caption{The convex lens attack on an iPhone with $d_{o1}=9m$, $f=50cm$, and various $d_b$.}
	\label{fig:cv_overview_9m_50cm_db}
\end{figure*}

For the second attack scenario in the convex lens attacks, the formed image is inverted from the attack lens. Besides, it also requires the $d_b$ to be as large as $f$. Due to the inverted image and large $d_b$, it is also impractical to apply this attack physically. First, the inverted image may be detected by the AD error detection system easily. Second, because the $d_b$ is so large, e.g., $25cm$, it is difficult to conceal the attack lens without being noticed by a human driver.

In terms of the third attack scenario, it works with various $f$ and $d_b$. More detailed discussions regarding the feasible attack types in the physical world are discussed in Appendix~\ref{sec:Attack Types in Real AD Scenarios}. Some attack examples on iPhone with $d_{o1}=9m$, $f=50cm$ and various $d_b$ are shown in Fig.~\ref{fig:cv_overview_9m_50cm_db}. The attacked region is quite blurry with the larger $d_b$ because the formed image is focused in front of the image sensor, which is similar to a near-eyesight case. However, when $d_b$ is smaller, the blurriness is released. For example, when $d_b=4cm$, the blurriness is reduced significantly. In terms of the disparity map, as the blurriness is reduced, the disparity map becomes clearer and more accurate. Meanwhile, the result indicates that a larger $d_b$ can introduce a more significant impact on the depth. Note that this attack can also work for different attack distances as long as $d_{o1}>f$. We can also notice that a larger $f$ has more effect on the depth from Table~\ref{tab:attack_error_rate_convex}.

\begin{table*}[]
\centering
\caption{The AER of the convex lens attack with the various object distance $d_{o1}$, $f$, and $d_b$.}
\label{tab:attack_error_rate_convex}
\resizebox{\textwidth}{!}{%
\begin{tabular}{|cc|ccccc|ccccc|ccccc|}
\hline
\multicolumn{2}{|c|}{} &
  \multicolumn{5}{c|}{$d_{o1}$=6m} &
  \multicolumn{5}{c|}{$d_{o1}$=9m} &
  \multicolumn{5}{c|}{$d_{o1}$=12m} \\ \cline{3-17} 
\multicolumn{2}{|c|}{\multirow{-2}{*}{Attack Parameters}} &
  \multicolumn{1}{c|}{} &
  \multicolumn{2}{c|}{\begin{tabular}[c]{@{}c@{}}Experimental Depth\\ (meters)\end{tabular}} &
  \multicolumn{2}{c|}{\begin{tabular}[c]{@{}c@{}}AER\\ (\%)\end{tabular}} &
  \multicolumn{1}{c|}{} &
  \multicolumn{2}{c|}{\begin{tabular}[c]{@{}c@{}}Experimental Depth\\ (meters)\end{tabular}} &
  \multicolumn{2}{c|}{\begin{tabular}[c]{@{}c@{}}AER\\ (\%)\end{tabular}} &
  \multicolumn{1}{c|}{} &
  \multicolumn{2}{c|}{\begin{tabular}[c]{@{}c@{}}Experimental Depth\\ (meters)\end{tabular}} &
  \multicolumn{2}{c|}{\begin{tabular}[c]{@{}c@{}}AER\\ (\%)\end{tabular}} \\ \cline{1-2} \cline{4-7} \cline{9-12} \cline{14-17} 
\multicolumn{1}{|c|}{$f$} &
  $d_b$ &
  \multicolumn{1}{c|}{\multirow{-2}{*}{\begin{tabular}[c]{@{}c@{}}Expected\\ Depth (meters)\end{tabular}}} &
  \multicolumn{1}{c|}{AV} &
  \multicolumn{1}{c|}{iPhone} &
  \multicolumn{1}{c|}{AV} &
  iPhone &
  \multicolumn{1}{c|}{\multirow{-2}{*}{\begin{tabular}[c]{@{}c@{}}Expected\\ Depth (meters)\end{tabular}}} &
  \multicolumn{1}{c|}{AV} &
  \multicolumn{1}{c|}{iPhone} &
  \multicolumn{1}{c|}{AV} &
  iPhone &
  \multicolumn{1}{c|}{\multirow{-2}{*}{\begin{tabular}[c]{@{}c@{}}Expected\\ Depth (meters)\end{tabular}}} &
  \multicolumn{1}{c|}{AV} &
  \multicolumn{1}{c|}{iPhone} &
  \multicolumn{1}{c|}{AV} &
  iPhone \\ \hline
\multicolumn{1}{|c|}{20cm} &
  2cm &
  \multicolumn{1}{c|}{4.67} &
  \multicolumn{1}{c|}{-} &
  \multicolumn{1}{c|}{4.84} &
  \multicolumn{1}{c|}{-} &
  3.62 &
  \multicolumn{1}{c|}{6.98} &
  \multicolumn{1}{c|}{-} &
  \multicolumn{1}{c|}{8.90} &
  \multicolumn{1}{c|}{-} &
  27.49 &
  \multicolumn{1}{c|}{9.29} &
  \multicolumn{1}{c|}{-} &
  \multicolumn{1}{c|}{8.39} &
  \multicolumn{1}{c|}{-} &
  9.74 \\ \hline
\multicolumn{1}{|c|}{20cm} &
  4cm &
  \multicolumn{1}{c|}{4.08} &
  \multicolumn{1}{c|}{6.08} &
  \multicolumn{1}{c|}{5.80} &
  \multicolumn{1}{c|}{49.17} &
  42.26 &
  \multicolumn{1}{c|}{6.09} &
  \multicolumn{1}{c|}{7.92} &
  \multicolumn{1}{c|}{9.66} &
  \multicolumn{1}{c|}{30.07} &
  58.66 &
  \multicolumn{1}{c|}{8.10} &
  \multicolumn{1}{c|}{12.07} &
  \multicolumn{1}{c|}{12.16} &
  \multicolumn{1}{c|}{49.10} &
  50.16 \\ \hline
\multicolumn{1}{|c|}{20cm} &
  8cm &
  \multicolumn{1}{c|}{2.90} &
  \multicolumn{1}{c|}{10.73} &
  \multicolumn{1}{c|}{6.88} &
  \multicolumn{1}{c|}{{\color[HTML]{FE0000} 269.86}} &
  137.27 &
  \multicolumn{1}{c|}{4.31} &
  \multicolumn{1}{c|}{5.27} &
  \multicolumn{1}{c|}{7.54} &
  \multicolumn{1}{c|}{22.31} &
  74.93 &
  \multicolumn{1}{c|}{5.72} &
  \multicolumn{1}{c|}{7.00} &
  \multicolumn{1}{c|}{12.31} &
  \multicolumn{1}{c|}{22.45} &
  115.18 \\ \hline
\multicolumn{1}{|c|}{20cm} &
  12cm &
  \multicolumn{1}{c|}{1.74} &
  \multicolumn{1}{c|}{15.59} &
  \multicolumn{1}{c|}{4.57} &
  \multicolumn{1}{c|}{{\color[HTML]{FE0000} 796.73}} &
  {\color[HTML]{FE0000} 162.69} &
  \multicolumn{1}{c|}{2.55} &
  \multicolumn{1}{c|}{5.30} &
  \multicolumn{1}{c|}{14.12} &
  \multicolumn{1}{c|}{{\color[HTML]{FE0000} 108.04}} &
  {\color[HTML]{FE0000} 453.99} &
  \multicolumn{1}{c|}{3.36} &
  \multicolumn{1}{c|}{7.62} &
  \multicolumn{1}{c|}{5.93} &
  \multicolumn{1}{c|}{{\color[HTML]{FE0000} 126.80}} &
  {\color[HTML]{FE0000} 76.40} \\ \hline
\multicolumn{1}{|c|}{30cm} &
  2cm &
  \multicolumn{1}{c|}{5.13} &
  \multicolumn{1}{c|}{-} &
  \multicolumn{1}{c|}{7.07} &
  \multicolumn{1}{c|}{-} &
  37.80 &
  \multicolumn{1}{c|}{7.67} &
  \multicolumn{1}{c|}{-} &
  \multicolumn{1}{c|}{8.96} &
  \multicolumn{1}{c|}{-} &
  16.81 &
  \multicolumn{1}{c|}{10.21} &
  \multicolumn{1}{c|}{-} &
  \multicolumn{1}{c|}{10.90} &
  \multicolumn{1}{c|}{-} &
  6.75 \\ \hline
\multicolumn{1}{|c|}{30cm} &
  4cm &
  \multicolumn{1}{c|}{4.73} &
  \multicolumn{1}{c|}{4.91} &
  \multicolumn{1}{c|}{7.35} &
  \multicolumn{1}{c|}{3.79} &
  55.17 &
  \multicolumn{1}{c|}{7.07} &
  \multicolumn{1}{c|}{8.19} &
  \multicolumn{1}{c|}{10.17} &
  \multicolumn{1}{c|}{15.79} &
  43.75 &
  \multicolumn{1}{c|}{9.42} &
  \multicolumn{1}{c|}{10.36} &
  \multicolumn{1}{c|}{14.52} &
  \multicolumn{1}{c|}{10.01} &
  54.19 \\ \hline
\multicolumn{1}{|c|}{30cm} &
  8cm &
  \multicolumn{1}{c|}{3.95} &
  \multicolumn{1}{c|}{4.97} &
  \multicolumn{1}{c|}{6.22} &
  \multicolumn{1}{c|}{{\color[HTML]{333333} 25.92}} &
  57.40 &
  \multicolumn{1}{c|}{5.89} &
  \multicolumn{1}{c|}{7.95} &
  \multicolumn{1}{c|}{7.50} &
  \multicolumn{1}{c|}{34.99} &
  27.36 &
  \multicolumn{1}{c|}{7.83} &
  \multicolumn{1}{c|}{6.25} &
  \multicolumn{1}{c|}{11.42} &
  \multicolumn{1}{c|}{20.17} &
  45.89 \\ \hline
\multicolumn{1}{|c|}{30cm} &
  12cm &
  \multicolumn{1}{c|}{3.18} &
  \multicolumn{1}{c|}{15.42} &
  \multicolumn{1}{c|}{5.58} &
  \multicolumn{1}{c|}{{\color[HTML]{FE0000} 385.54}} &
  {\color[HTML]{FE0000} 75.54} &
  \multicolumn{1}{c|}{4.72} &
  \multicolumn{1}{c|}{6.19} &
  \multicolumn{1}{c|}{7.99} &
  \multicolumn{1}{c|}{31.18} &
  69.35 &
  \multicolumn{1}{c|}{6.26} &
  \multicolumn{1}{c|}{6.17} &
  \multicolumn{1}{c|}{9.39} &
  \multicolumn{1}{c|}{1.45} &
  50.07 \\ \hline
\multicolumn{1}{|c|}{50cm} &
  2cm &
  \multicolumn{1}{c|}{5.50} &
  \multicolumn{1}{c|}{-} &
  \multicolumn{1}{c|}{6.86} &
  \multicolumn{1}{c|}{-} &
  24.70 &
  \multicolumn{1}{c|}{8.22} &
  \multicolumn{1}{c|}{-} &
  \multicolumn{1}{c|}{9.03} &
  \multicolumn{1}{c|}{-} &
  9.82 &
  \multicolumn{1}{c|}{10.95} &
  \multicolumn{1}{c|}{-} &
  \multicolumn{1}{c|}{11.51} &
  \multicolumn{1}{c|}{-} &
  5.11 \\ \hline
\multicolumn{1}{|c|}{50cm} &
  4cm &
  \multicolumn{1}{c|}{5.26} &
  \multicolumn{1}{c|}{5.73} &
  \multicolumn{1}{c|}{7.18} &
  \multicolumn{1}{c|}{{\color[HTML]{333333} 8.90}} &
  36.47 &
  \multicolumn{1}{c|}{7.87} &
  \multicolumn{1}{c|}{8.68} &
  \multicolumn{1}{c|}{11.63} &
  \multicolumn{1}{c|}{10.38} &
  47.88 &
  \multicolumn{1}{c|}{10.47} &
  \multicolumn{1}{c|}{11.75} &
  \multicolumn{1}{c|}{14.92} &
  \multicolumn{1}{c|}{12.26} &
  42.54 \\ \hline
\multicolumn{1}{|c|}{50cm} &
  8cm &
  \multicolumn{1}{c|}{4.79} &
  \multicolumn{1}{c|}{4.99} &
  \multicolumn{1}{c|}{5.80} &
  \multicolumn{1}{c|}{{\color[HTML]{333333} 4.19}} &
  21.11 &
  \multicolumn{1}{c|}{7.16} &
  \multicolumn{1}{c|}{8.18} &
  \multicolumn{1}{c|}{8.16} &
  \multicolumn{1}{c|}{14.35} &
  14.08 &
  \multicolumn{1}{c|}{9.52} &
  \multicolumn{1}{c|}{10.72} &
  \multicolumn{1}{c|}{10.75} &
  \multicolumn{1}{c|}{12.60} &
  12.94 \\ \hline
\multicolumn{1}{|c|}{50cm} &
  12cm &
  \multicolumn{1}{c|}{4.33} &
  \multicolumn{1}{c|}{7.57} &
  \multicolumn{1}{c|}{6.79} &
  \multicolumn{1}{c|}{{\color[HTML]{FE0000} 75.04}} &
  56.99 &
  \multicolumn{1}{c|}{6.45} &
  \multicolumn{1}{c|}{7.90} &
  \multicolumn{1}{c|}{7.43} &
  \multicolumn{1}{c|}{22.43} &
  15.13 &
  \multicolumn{1}{c|}{8.57} &
  \multicolumn{1}{c|}{8.03} &
  \multicolumn{1}{c|}{9.87} &
  \multicolumn{1}{c|}{6.34} &
  15.06 \\ \hline
\multicolumn{1}{|c|}{None} &
  None &
  \multicolumn{1}{c|}{6} &
  \multicolumn{1}{c|}{6.50} &
  \multicolumn{1}{c|}{6.89} &
  \multicolumn{1}{c|}{8.41} &
  14.78 &
  \multicolumn{1}{c|}{9} &
  \multicolumn{1}{c|}{8.60} &
  \multicolumn{1}{c|}{9.94} &
  \multicolumn{1}{c|}{4.49} &
  10.40 &
  \multicolumn{1}{c|}{12} &
  \multicolumn{1}{c|}{11.67} &
  \multicolumn{1}{c|}{12.67} &
  \multicolumn{1}{c|}{2.78} &
  5.59 \\ \hline
\multicolumn{2}{|c|}{\textbf{Average AER}} &
  \multicolumn{3}{c|}{\textbf{}} &
  \multicolumn{1}{c|}{\textbf{18.39}} &
  \textbf{47.28} &
  \multicolumn{3}{c|}{\textbf{}} &
  \multicolumn{1}{c|}{\textbf{22.69}} &
  \textbf{36.84} &
  \multicolumn{3}{c|}{\textbf{}} &
  \multicolumn{1}{c|}{\textbf{16.80}} &
  \textbf{37.06} \\ \hline
\end{tabular}%
}
\end{table*}

Furthermore, to investigate the AER in the physical world, we calculate the expected depth in meters using Eqs.~(\ref{equ:magnification}) and (\ref{equ:convex}). Table~\ref{tab:attack_error_rate_convex} shows the AER of the convex lens attack with varied $f$ and $d_b$ values for different object distances. Similar to the results of the concave lens attacks in Table~\ref{tab:attack_error_rate_concave}, when $d_b$ is $2cm$, the AER is around 14\%.

\begin{figure}[t]
\centering
	\includegraphics[width=0.8\columnwidth]{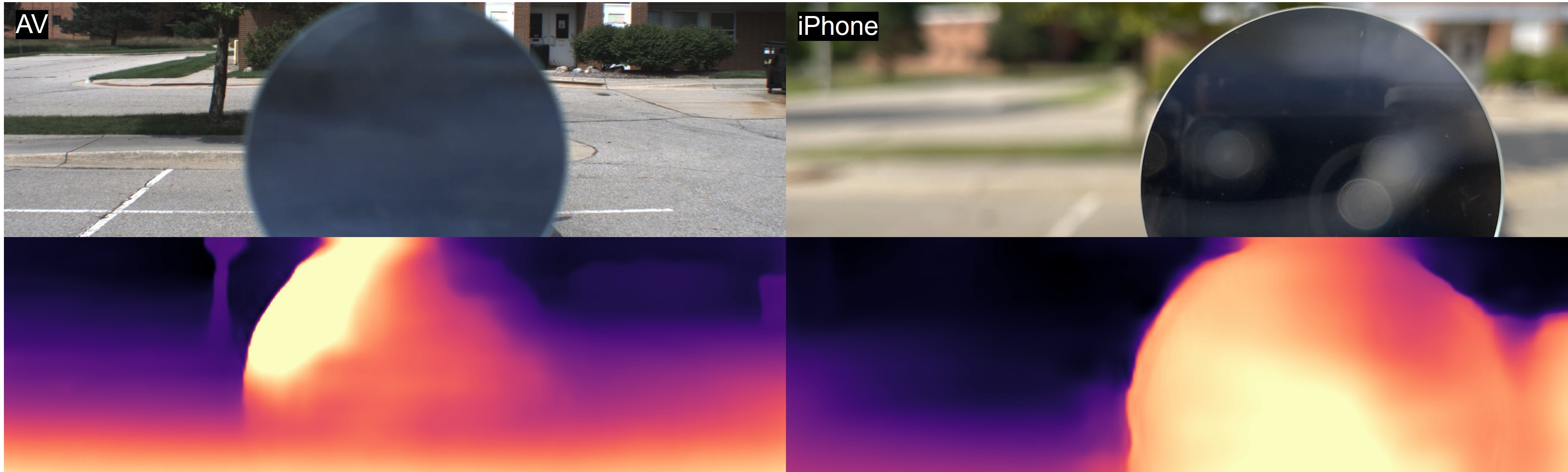}
	\caption{The convex lens attack on AV and iPhone cameras with $d_{o1}=6m$, $f=20cm$, and $d_b=12cm$.}
	\label{fig:non_success_convex}
\end{figure} 

However, for partial image attacks, with the smaller $f$ and larger $d_b$, the AER increases. Sometimes, it could increase to more than 100\%. 
We show an example in Figure~\ref{fig:non_success_convex}.
When we look at the convex lens attack on AV and iPhone cameras with $d_{o1}=6m$, $f=20cm$, and $d_b=12cm$, we can see that it is much more blurry in the in-lens or out-of-lens area compared to the case with the smaller $d_{o1}$ and larger $f$.
This occurs due to the introduction of a stronger depth of field, which is achieved by employing larger $d_{o1}$ values and smaller $f$ values. Due to the strong blurriness, convex lens attacks also obscure the target object, leading to detection failures in YOLO v8. Therefore, we mark the object detection failure case using red color in Table~\ref{tab:attack_error_rate_convex}. We notice that car detection failure often matches with a high AER, which is not preferred when the attacker launches the attack. Therefore, we calculate the average AER in the last row by omitting these red numbers. The overall AER on both AV and iPhone is much higher than the AER in concave lens attacks. Meanwhile, the average AER on the iPhone is usually higher than that on AV. The average AER across both the AV and iPhone is 29.84\%.

\cbox{To summarize, \attack works for both AV and iPhone in the physical world in partial and full image attacks. The overall AER is lower on iPhone in the concave lens attack and lower on AV in the convex lens attack with approximately 11.48\% and 29.84\%, respectively.
It demonstrates the effectiveness of \attack in real AD scenarios. 
}

\section{Discussion and Limitation}\label{sec:discussion}

We discuss some common concerns of our attack regarding attack types in real AD scenarios, attack on AV with multi-camera, the generality, and the limitation. We also present potential defense methods against \attack. 

\textbf{Attack Types in Real AD Scenarios.}
In the physical attack, regarding the practicality and stealthiness of the attack,  $d_b$ is usually not large, e.g., $5cm$. Besides, the object is often far from the camera, e.g., $10m$. Because of the constraints, only the concave lens attack and the third attack scenario of the convex lens attack are feasible. Therefore, as discussed in Section~\ref{sec:Real-World Physical Attack}, we  demonstrate these two attacks in physical AD scenarios. More detailed mathematical analysis can be found in Appendix~\ref{sec:Attack Types in Real AD Scenarios}.

\textbf{Attack on AV with Multi-camera.}
Commercial AVs, like Tesla, usually are equipped with more than one camera to sense the environment~\cite{tesla_camera}. Popular Tesla Vision~\cite{tesla_vision} based models, such as Model 3 and Model Y, are equipped with eight cameras and powerful vision processing which can stitch the image together and provide 360 degrees of visibility at up to 250 meters of range~\cite{tesla_camera,tesla_Stitching}. \attack can also work on AV when it is equipped with multiple cameras. To verify our attack on the AV, we apply the attack on the front middle camera, and then collect the 3 different front camera images. We stitch the images to show that the attack is not affected by the multiple cameras from the stitched image and the disparity map. 

\begin{figure}[t]
\centering
	\includegraphics[width=0.65\columnwidth]{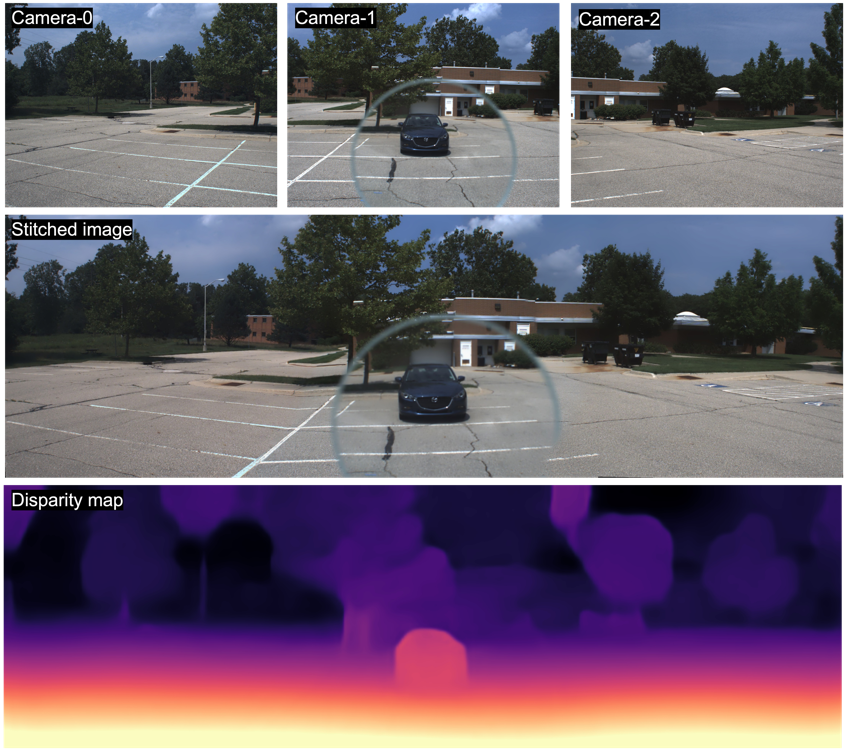}
	\caption{The images in the first row are captured from the 3 cameras on the AV, which are used to form the stitched image. The last image shows the disparity map of the stitched image.}
	\label{fig:stitched_image_depth}
\end{figure}

An example is shown in Figure~\ref{fig:stitched_image_depth}, we collect images from the 3 FILR cameras on AV, i.e., camera-0, camera-1, and camera-2. Each of the cameras is with a $60\degree$ field of view horizontally. To emulate the vision processing on Tesla, we perform image stitching using these camera images. Then, we obtain its disparity map from Monodepth2. It can be seen that the attack is not affected by the multiple cameras from the stitched image and the disparity map. Therefore, \attack would still work for multi-camera AVs.

\textbf{Generality.}
The majority of car camera hardware could be vulnerable to \attack. 
By adjusting $f$ and $d_b$, we can manipulate the object depth. 
As the focal lengths of most car cameras are similar, by adjusting $d_b$, they will all fall victim to the proposed attack. Meanwhile, since the camera image serves as the input for the MDE models, the output depth map will be affected by our attack regardless of the model variations. 

Regarding vehicle software, existing AD systems have advanced object detection models. Despite being able to identify the presence of objects ahead in our experiments, these models lack distance information, making them ineffective in preventing our attack.

\textbf{Limitation.}
The primary limitation lies in the stealthiness of our attack. Captured images/video streams from the camera sensors are usually not visible to the human driver. However, our current attack prototype is noticeable. For commercial AVs, autopilot cameras are usually in the cabin or the car body. 
The car manufacturer usually installs the camera in an unobtrusive place by shielding it with the vehicle corners or the vehicle decoration~\cite{tesla_exterior}. 
The autopilot camera sensors are usually tiny, which allows us to launch our attack by simply sticking, taping, or absorbing the attack device on the car body near the car camera. We simulate our attack device implementation on Tesla Model 3 in Appendix~\ref{app:manufactured}.
To make the attack more stealthy, we could design and print a 
3D lens holder using some transparent materials, such as glass or plastic. However, stealthiness can still become an issue if the designed attack device is too large or placed in a relatively conspicuous place. 

Second, in the presence of a highly alert AD system, our attack could potentially be thwarted by blur detection. This could trigger an alert and prompt the system to cease operation until the attack lens is removed from the camera sensor. Therefore, blur detection can serve as a defense mechanism. We elaborate on various blur detection methods in the defense section below.

The final limitation is related to changing $f$ and controlling $d_b$. We know that $f$ and $d_b$ are the key parameters in controlling the attack depth. However, it is not easy to change the focal length of the attack lens $f$ without accessing the victim's vehicle. Therefore, we control $d_b$ using a design lens slider or a phone slider in the experiments to continuously manipulate the attack depth. With an increasing cost, a more sophisticated version of our attack could employ a purpose-built apparatus similar in principle to a consumer digital camera lens, perhaps even employing electrically tunable lenses to precisely calibrate focus, focal length, and lens distance.




\textbf{Defense.}
One of the most effective countermeasures is by using sensor fusion~\cite{wang2017sonic}, which combines the outputs of multiple sensors to produce an accurate result. Most AVs employ sensors other than cameras on their vehicles, such as lidar, radar, and ultrasonic sensors. Unlike monocular cameras, these sensors convey accurate depth information, which obviates the need for monocular depth detection methods, and thus can defeat our attack. However, depending on the architecture of the fusion method in question, if monocular camera depth information is still used, our method could still reduce the accuracy. Manufacturers should test their systems against \attack to ensure that the fusion method they use is sufficiently robust.

Another defense method is to add a detection module that detects image blur, either in-lens blur or out-of-lens blur, caused by the proposed attack. 
Once a blur in the image is detected, the AV should trigger an alert, prompt the system to cease operation, and warn the human driver to inspect the physical condition of the car camera. For example, there are some methods to detect the image blur: variation of the Laplacian (VarLap)~\cite{Adrian_Rosebrock,WillBrennan}, High-frequency multiscale Fusion and Sort Transform (HiFST)~\cite{alireza2017spatially}, and local-based defocus blur segmentation (LDB)~\cite{yi2016lbp}. All these methods can well distinguish between the blurred and non-blurred regions in the image, meaning that they can be potentially used to detect our proposed attack.

In addition to employing the aforementioned methods for detecting the attack, there is potential to devise an attack-aware depth estimation algorithm. This algorithm could correct the compromised depth by deblurring the image and rescaling the attacked area to the correct size and depth. We identify this as an area for future investigation and development.

\section{Related Work}

Zhang et al. are the first to investigate white-box adversarial attacks on MDE~\cite{zhang2020adversarial}. They employed imperceptible perturbations to execute three distinct types of attacks: non-targeted attacks on a specific image, targeted attacks on a particular object within an image, and universal attacks that can be applied to any image. Similarly, Wong et al. present a method for using imperceptible additive adversarial perturbations to selectively alter the perceived geometry of a scene for MDE~\cite{wong2020targeted}. To generalize the attacks, Daimo et al. propose black-box adversarial attacks on MDE using evolutionary multi-objective optimization~\cite{daimo2021black}. All of these attacks, however, are confined to the digital world and lack real-world applicability.

Recently, Yamanaka et al. devise artificial adversarial patches capable of deceiving the target methods into providing inaccurate depth estimations for the regions where these patterns were applied~\cite{yamanaka2020adversarial}. Different from the noticeable attack patches, Cheng et al. employ an optimization-based method to generate inconspicuous adversarial patches that are tailored towards physical objects to attack depth estimation~\cite{cheng2022physical}. Compared with existing white-box physical attacks, \attack enables a new genre of physical attack using optical lenses in a black-box setting, which is even more general and more robust.

\section{Conclusion}
In this paper, we present \attack, a new genre of physical attack towards MDE based AD systems using optical lenses.
By exploring the vulnerability of MDE, we formulate concave lens attack and convex lens attack mathematically. 
We conduct the attack simulation with different MDE algorithms to showcase the feasibility of our attack. Through extensive real-world experiments, we find \attack is effective across diverse attack parameter settings and at various object distances. 
The successful demonstration of \attack on monocular vision-based depth estimation suggests potential security implications on real-world AD systems.

\section*{Acknowledgements}

We would like to extend our appreciation to the anonymous reviewers for their invaluable input on our
study. This work was supported in part by the U.S. National
Science Foundation grant CNS-2235231.

\bibliographystyle{splncs04}
\bibliography{Reference}

\appendix
\section{Appendix}
\subsection{Pinhole Camera Model}\label{sec:Pinhole Camera Model}

\begin{figure}
\centering
	\includegraphics[width=0.5\textwidth]{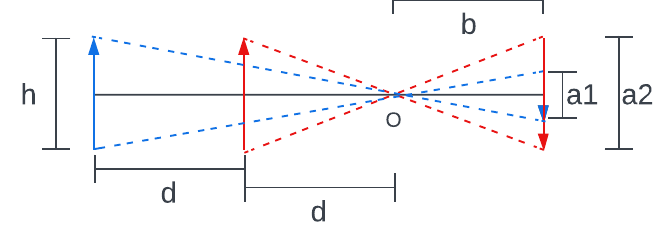}
	\caption{The perspective projection of the pinhole camera.}
	\label{fig:pinhole}
\end{figure}

The perspective projection of the pinhole camera is shown in Fig.~\ref{fig:pinhole}. $O$ denotes the pinhole position. $h$ represents the size of the two objects. $b$ is the distance between the pinhole to the image sensor, and $d$ and $2d$ stand for the distance between the two same-size objects and the pinhole, respectively. Based on the triangulation, we have: 
\begin{equation}\label{pinholeequ1}
    \frac{h}{d}=\frac{a_{1}}{b},
\end{equation}
\begin{equation}\label{pinholeequ2}
    \frac{h}{2d}=\frac{a_{2}}{b}.
\end{equation}    
Now, Using Eqs.~(\ref{pinholeequ1}) and (\ref{pinholeequ2}), we have: 
\begin{equation}
    \frac{2d}{d}=\frac{a_{1}}{a_{2}}=2.
\end{equation}   
Thus, we can conclude that the depth of the same object is inversely proportional to its object size in the image.

\subsection{Attack Device on Manufactured Vehicle}\label{app:manufactured}

\begin{figure}[t]
    \centering    
    \subfigure[Full view of the attack]{\includegraphics[width=0.4\textwidth]{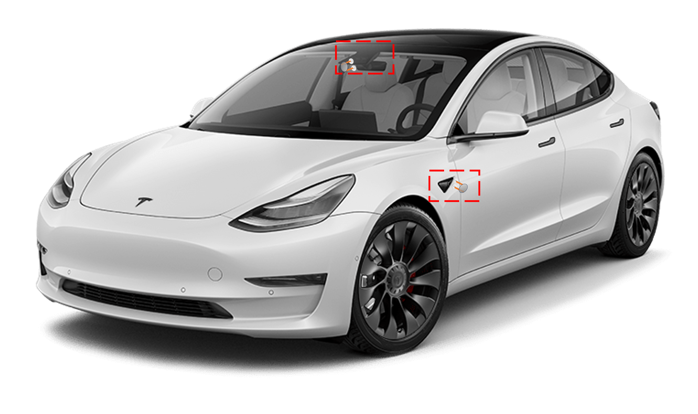}}
    \subfigure[Attack device on front camera]{\includegraphics[width=0.28\textwidth]{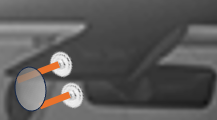}} 
    \subfigure[Attack device on side camera]{\includegraphics[width=0.28\textwidth]{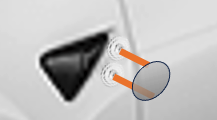}}      
    \caption{(a) In deploying the attack device on the Tesla Model 3, the designated location for the autopilot cameras is indicated by the red dotted line. The attack device consists of an attack lens (greyish transparent ellipse), a lens holder (the two yellowish sticks), two fixed suction cups (the bluish dots), and a remote control module (not shown in the figure). (b) and (c) shows the attack device implemented on the front and side cameras in detail. }
    \label{fig:Manufactured}
\end{figure}

We simulate the attack device on a manufactured vehicle (i.e., Tesla Model 3) in Fig.~\ref{fig:Manufactured}. The attack device comprises an attack lens, a lens holder, two fixed suction cups, and a remote control module (ensuring the proposed intermittent attack). Using two fixed suction cups, the attack device can be placed in different locations. Compromising the front autopilot camera may result in a collision with the car in front or behind. Conversely, an attack on the side autopilot camera can lead to a collision with another car in an adjacent lane when the AV is changing lanes.

Concerning stealthiness, from Fig.~\ref{fig:experimental_setup}(a) and Fig.~\ref{fig:Manufactured}(a), it is evident that the attack device appears relatively small in the overall view of the attack, making it less likely to be noticed by the human driver if they are not actively paying attention. However, when the attack targets the front camera, it becomes more noticeable to the human driver. In contrast, the attack is more stealthy when directed at the side camera.

\subsection{Attack Types in Real AD Scenarios}\label{sec:Attack Types in Real AD Scenarios}

In real AD scenarios, the value of $d_b$ typically remains relatively small, for instance, around $5cm$. Additionally, the object is frequently situated at a considerable distance from the camera, often reaching approximately $10m$. In a concave lens attack, no matter what the value of $d_{o1}$ is, the formed image is always virtual and upright, meaning that it always works.

Regarding the convex lens attack, when $0<d_{o1}<f$ (first attack scenario), the formed images are real and upright in the combination lenses as long as $d_b \geq 0$. However, the object will be very close to the camera (e.g., $50cm$), which is not normal in AD scenarios. For the second attack scenario, when $d_{o1}>f$, the formed images will either become real and inverted when $d_b-|d_{i1}|>f_c$, or virtual and upright when $0<d_b-|d_{i1}|<f_c$. Since $f_c$ is a very small value, we need to ensure $d_b>|d_{i1}|$, meaning that the distance between the attack lens and camera is larger than the focal length of the attack lens $f$, which contradicts the physical attack scenarios as discussed in the Section~\ref{sec:Real-World Physical Attack}. For the third attack scenario, $d_{o1}>f$ and $d_b<|d_{i1}|$, the formed images are real and upright.
Here, we highlight that only the concave lens attack and the third attack scenario of the convex lens attack are feasible in practical AD scenarios.

\end{document}